\documentstyle[psfig]{mn2e}

\begin{document}

\title[Collisional Processes in Extrasolar Planetesimal Disks]
  {Collisional Processes in Extrasolar Planetesimal Disks --- \\
   Dust Clumps in Fomalhaut's Debris Disk}

\author[M. C. Wyatt and W. R. F. Dent]
  {M. C. Wyatt\thanks{Email: wyatt@roe.ac.uk}
  and W. R. F. Dent\\
  UK Astronomy Technology Centre, Royal Observatory,
  Blackford Hill, Edinburgh EH9 3HJ, UK}

\maketitle

\begin{abstract}
This paper presents a model for the outcome of collisions between
planetesimals in a debris disk and assesses the impact of collisional
processes on the structure and size distribution of the disk.
The model is presented by its application to Fomalhaut's collisionally
replenished dust disk;
a recent 450 $\mu$m image of this disk shows a clump embedded within it
with a flux $\sim 5$ per cent of the total.
The following conclusions are drawn:
\textbf{(i)} SED modelling is consistent with Fomalhaut's disk having a
collisional cascade size distribution extending from bodies 0.2 m
in diameter (the largest that contribute to the 850 $\mu$m flux)
down to 7 $\mu$m-sized dust (smaller grains are blown out of the
system by radiation pressure).
\textbf{(ii)} Collisional lifetime arguments imply that the collisional
cascade starts with planetesimals 1.5--4 km in diameter, and so has a mass
of 20--30 $M_\oplus$.
Any larger bodies must be predominantly primordial.
\textbf{(iii)} Constraints on the timescale for the ignition of the
collisional cascade from planet formation models are consistent with
these primordial planetesimals having the same distribution as the
cascade extending up to 1000 km, resulting in a disk mass of 5--10 times
the minimum mass solar nebula.
\textbf{(iv)} The debris disk is expected to be intrinsically clumpy,
since planetesimal collisions result in dust clumps that can last up to
700 orbital periods.
The intrinsic clumpiness of Fomalhaut's disk is below current detection
limits, but could be detectable by future observatories such as the
\textit{ALMA}, and could provide the only way of determining this
primordial planetesimal population.
Also, we note that such intrinsic clumpiness in an exozodiacal cloud-like
disk could present a confusion limit when trying to detect terrestrial
planets.
\textbf{(v)} The observed clump could have originated in a collision between
two runaway planetesimals, both larger than 1400 km diameter.
It appears unlikely that we should witness such an event unless both the
formation of these runaways and the ignition of the collisional cascade
occurred relatively recently (within the last $\sim 10$ Myr), however
this is a topic which would benefit from further exploration using planet
formation and collisional models.
\textbf{(vi)} Another explanation for Fomalhaut's clump is that
$\sim 5$ per cent of the planetesimals in the ring were trapped in
1:2 resonance with a planet orbiting at 80 AU when it migrated out due to
the clearing of a residual planetesimal disk.
The motion on the sky of such a clump would be 0.2 arcsec/year, and it
would be more prominent at shorter wavelengths.
\end{abstract}


\section{Introduction}
\label{sec-intro}
At least 15 per cent of main sequence stars exhibit detectable far-IR
emission in excess of that expected from the photosphere
(Plets \& Vynckier 1999; Lagrange, Backman \& Artymowicz 2001).
The spectral energy distribution (SED) of this excess implies that it is
thermal emission from dust in regions analogous to the Kuiper belt
(i.e, at $>30$ AU from the star).
The few stars for which this emission has been imaged confirm this
location for the dust and show that it is confined to narrow ring-like
disks (e.g., Holland et al.~1998; Greaves et al.~1998; Jayawardhana et
al.~1998).
The short lifetime inferred for this dust, due to both mutual collisions
and radiation forces, implies that it must be continually replenished,
and it is thought that the dust disk is fed by the collisional
grinding down of a population of larger planetesimals which may
have formed during the system's planetary formation phase
(Backman \& Paresce 1993).
Furthermore it has often been speculated that the absence of dust
close to the star in these systems has been caused by clearing of
the residual planetesimals by a planetary system.

The debris disk images also show that the dust rings are neither
smooth nor symmetrical:
recent sub-mm images of the Fomalhaut disk show a clump containing
5 per cent of the total disk flux (Holland et al.~2002);
there are several clumps in the $\epsilon$ Eridani disk, the largest
of which contains 6--7 per cent of the disk flux (Greaves et al.~1998);
a 5 per cent brightness asymmetry is seen in the structure of the
HR 4796 disk (Telesco et al.~2000), which could be accounted for by
a clump containing $\sim 0.5$ per cent of the disk flux;
Vega's dust disk is dominated by emission from two dust clumps
(Holland et al.~1998; Koerner, Sargent \& Ostroff 2001);
and $\beta$ Pictoris exhibits several small clumps in the mid-plane
of the NE extension of its optical disk (Kalas et al.~2000),
as well as a sub-mm emission peak lying $\sim$ 34 arcsec
SW of the star (Holland et al.~1998; Zuckerman 2001).

While clumps are now known to be a common feature of debris disks,
their origin, and whether this is the same in all cases, is still
unclear.
One way to form clumps is the gravitational perturbations
of unseen planets, since we know that their effect on the orbits of
planetesimals in these disks could cause the observed dust rings to
contain both offset and warp asymmetries (Wyatt et al.~1999; hereafter
WDT99) as well as resonant clumps (Wyatt 1999; Ozernoy et al.~2000).
Indeed the observed asymmetries have been modelled by various
authors and shown to be possible evidence of planets of mass between
$10M_\oplus$ and $2M_J$ orbiting close to the inner edge of the
rings (typically $\sim 30$ AU).
However, it is difficult to test this hypothesis as the proposed
planets are not detectable with current techniques, and such
interpretations must be treated with caution until alternative
explanations have been explored.

Another mechanism for forming clumps in these disks is stochastic
collisions between their largest planetesimals.
A collisionally produced disk is inherently clumpy because
collisional debris follows the orbit of the parent planetesimal until
it has had a chance to precess around that orbit.
The question is, do collisions of a magnitude large enough to produce an 
observable dust clump happen often enough, and the resulting clumps
last long enough, to make a disk appear clumpy?
The possibility of observing collisional clumps was first discussed
by Stern (1996) who modelled the expected clumpiness of our own
Kuiper belt, but the application to extrasolar debris disks has received
little attention, as it was thought that collisions large enough to produce
observable dust clumps would occur too infrequently, even if large enough
planetesimals do exist in the disks.
However, recent modelling of the asteroid belt stressed the importance
of stochastic collisions by showing that the collisional destruction of
asteroids large enough to more than double the brightness of the asteroid
belt should occur about once every 20 Myr (Durda \& Dermott 1997;
see fig.~19 of Grogan, Dermott \& Durda 2001).
This possibility may be supported by evidence of stochastic increases
in the interplanetary dust flux from $^3$He in deep-sea sediments
(Farley 1995).
Also, dust clumps caused by collisions in the asteroid belt may already
have been observed --- the temporary dust tail activity of the
asteroid-comet Elst/Pizarro may have resulted from collisions with
other main belt asteroids (Toth 2000).
This prompted Wyatt et al.~(2000) to revisit the collisional hypothesis
by applying a simple collisional model to the recently obtained image of
the Fomalhaut disk (Holland et al.~2002).
This implied that collisions were unlikely to be causing Fomalhaut's clump,
but they concluded that this possibility could not be ruled out because of
uncertainties in the disk's size distribution.

It is the purpose of this paper to explore in depth the role of
collisions in shaping the structure of the Fomalhaut disk.
In particular the aim is to determine whether the clump observed
by Holland et al.~(2002) could be collisional in origin.
One reason why it is crucial to test the collisional hypothesis is that
the only observation that has been proposed to test whether the
dust clumps are associated with planetary resonances relies on the
fact that such clumps would orbit the star with the planet ---
motion that should be detectable on timescales of a few years
(Ozernoy et al. 2000).
However, it may not be possible to use detection of such motion as an
unambiguous test for the presence of planets, since clumps that are
collisional in origin would also orbit the star.
After summarizing the observations of the Fomalhaut disk (section 2)
we consider the size distribution expected from theoretical arguments
(section 3), then estimate this distribution observationally by modelling
the observed SED (section 4).
A model of collisions is then developed which is used to explore the
role of collisional destruction in forming the observed size distribution
(section 5).
This model is then expanded to determine what magnitude of collision we
expect to see in Fomalhaut's disk due to the break-up of its planetesimals
(i.e., its intrinsic clumpiness; section 6).
Finally we explore the possible scenarios in which collisions might have
produced Fomalhaut's dust clump and propose another possible origin
for the clump (section 7).
The conclusions are given in section 8.

\section{The Fomalhaut Disk}
\label{sec-fom}
Fomalhaut was one of the first four main sequence stars found
to exhibit infrared emission in excess of that from the photosphere
and one of only a handful that have had that emission resolved.
It is an A3V star at a distance of $R_\star = 7.7$ pc that is estimated to
have an age of $t_\rmn{sys} = 200 \pm 100$ Myr based on both
evolutionary isochrone models (Lachaume et al.~1999; Song et al.~2001)
and on the ages of the Castor moving group with which it is associated
(Barrado Y Navascues 1998) and that of its common proper motion companion
GL879 (Barrado Y Navascues et al.~1997).
The stellar parameters that have been adopted in this work are:
$T_\rmn{eff} = 9060$ K, $\log{g}=4.29$, $M = 2M_\odot$ (Song et al.~2001);
and $L = 13L_\odot$ (Backman \& Paresce 1993).
In all modelling, the stellar spectrum is assumed to be that of a Kurucz
model atmosphere with solar metallicity and the above parameters.
No gas disk has been observed around Fomalhaut (Liseau 1999).

\begin{figure}
  \begin{center}
    \begin{tabular}{c}
      \psfig{figure=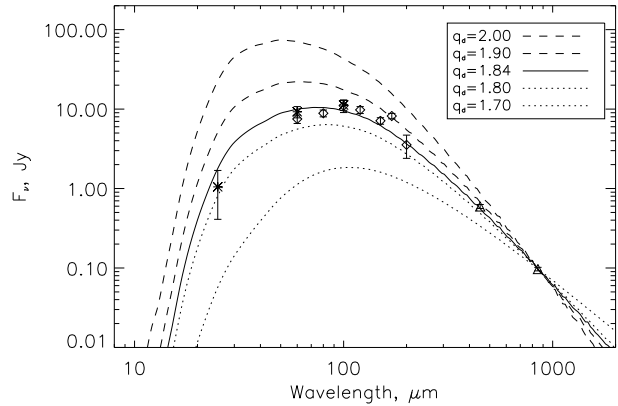,height=2.1in}
    \end{tabular}
    \caption{The SED of emission from Fomalhaut after subtraction of the
    stellar photosphere (see Table \ref{tab:fomsed}).
    The asterisks, diamonds and triangles correspond to the \textit{IRAS},
    \textit{ISO} and SCUBA fluxes, respectively.
    The lines represent model fits to these data (see section 4) that assume
    that the disk's spatial distribution is defined by a fit to the 450 $\mu$m
    image (Holland et al.~2002), and that it is comprised of solid grains
    of 1/3 (by volume) amorphous silicate and 2/3 organic refractory material
    (e.g., Li \& Greenberg 1997, 1998).
    The grains' size distributions in the models extend down to the radiation
    pressure blow-out limit, $D_\rmn{min} = 7$ $\mu$m, with power law indices
    given by $q_\rmn{d}$.}
    \label{fig:fomsed}
  \end{center}
\end{figure}

\begin{figure}
  \begin{center}
    \vspace{0.1in}
    \begin{tabular}{rr}
      \textbf{(a)} & \psfig{figure=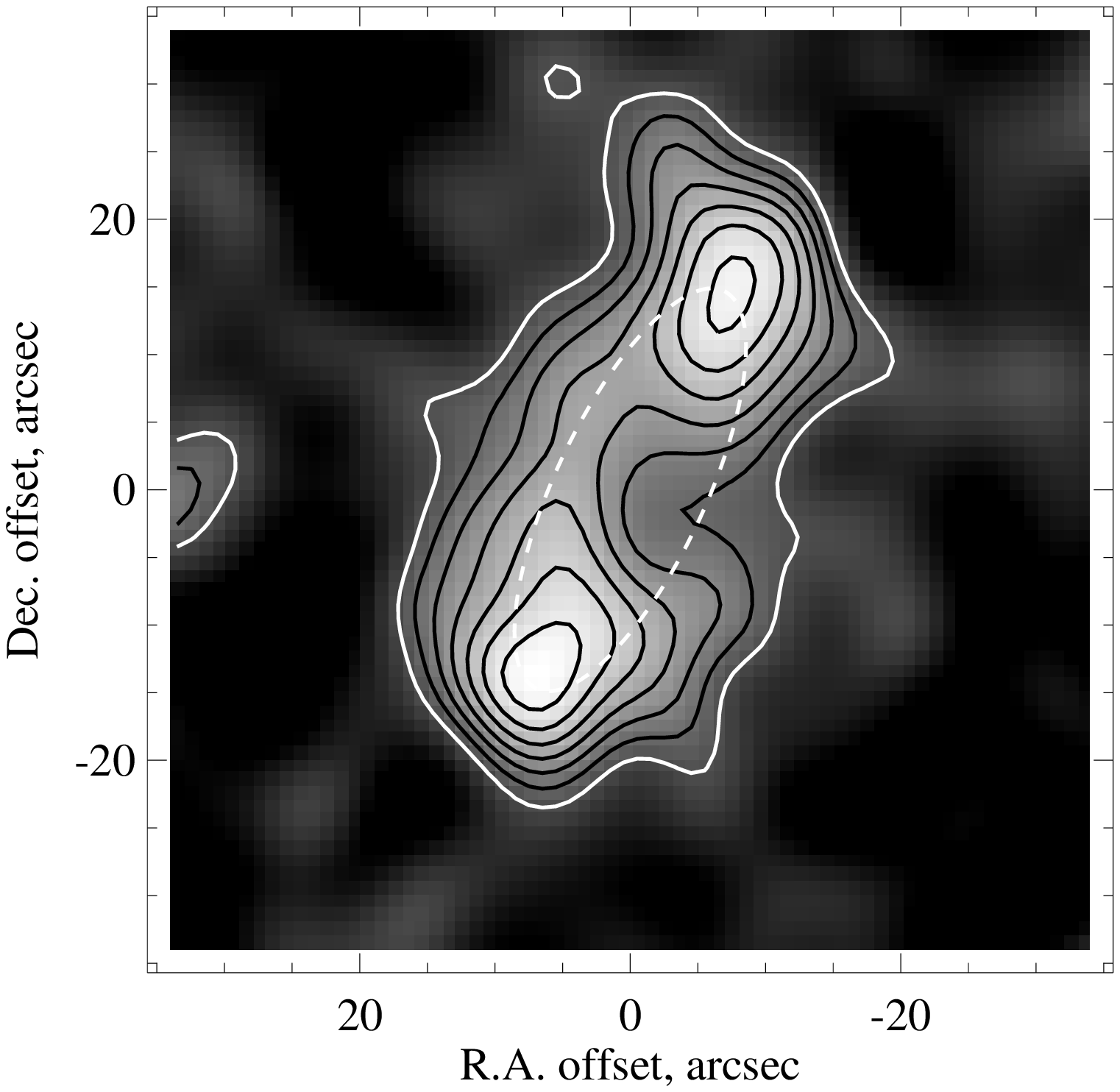,height=2.2in}   \\[0.15in]
      \textbf{(b)} & \psfig{figure=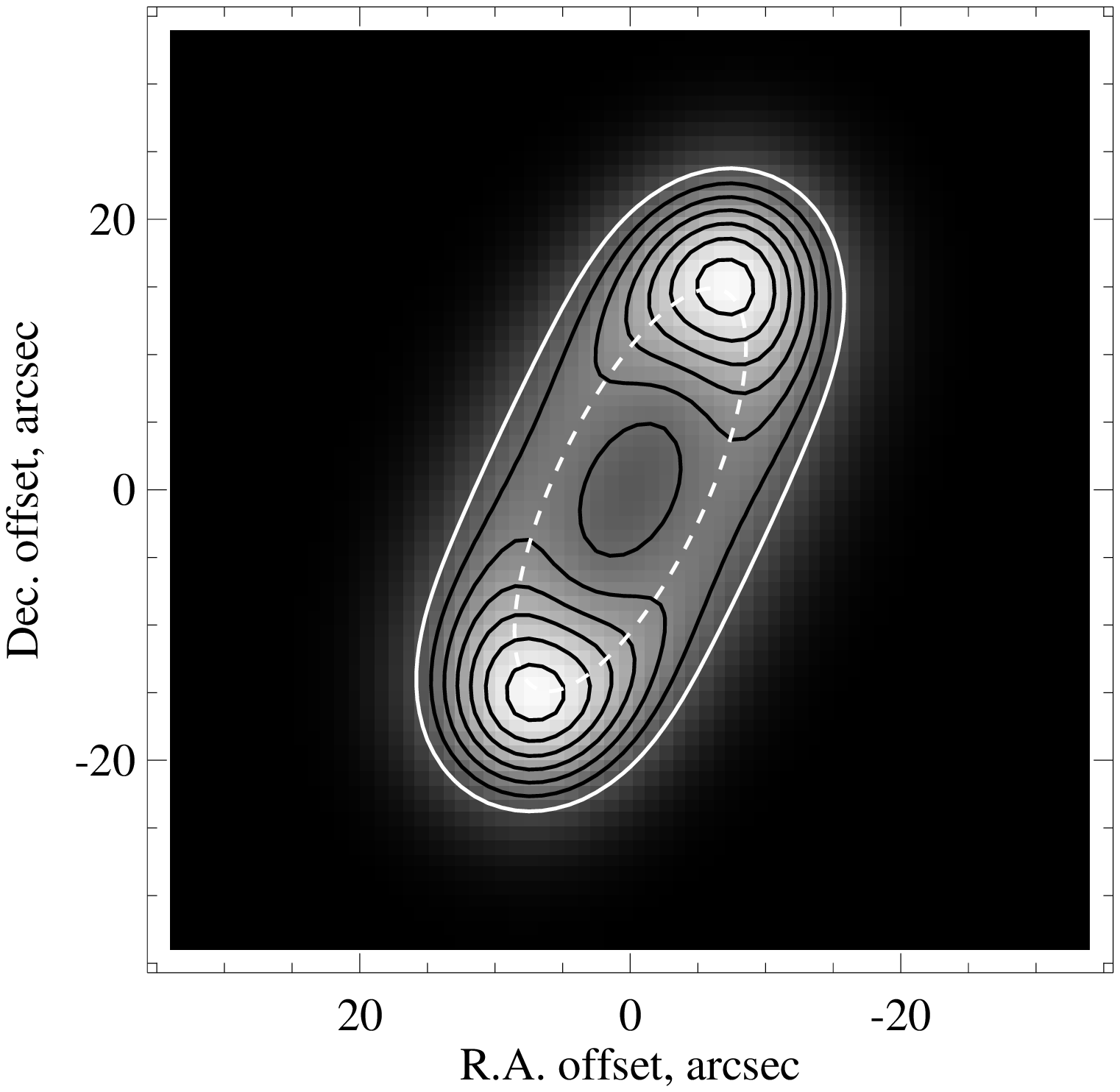,height=2.2in} \\[0.15in]
      \textbf{(c)} & \psfig{figure=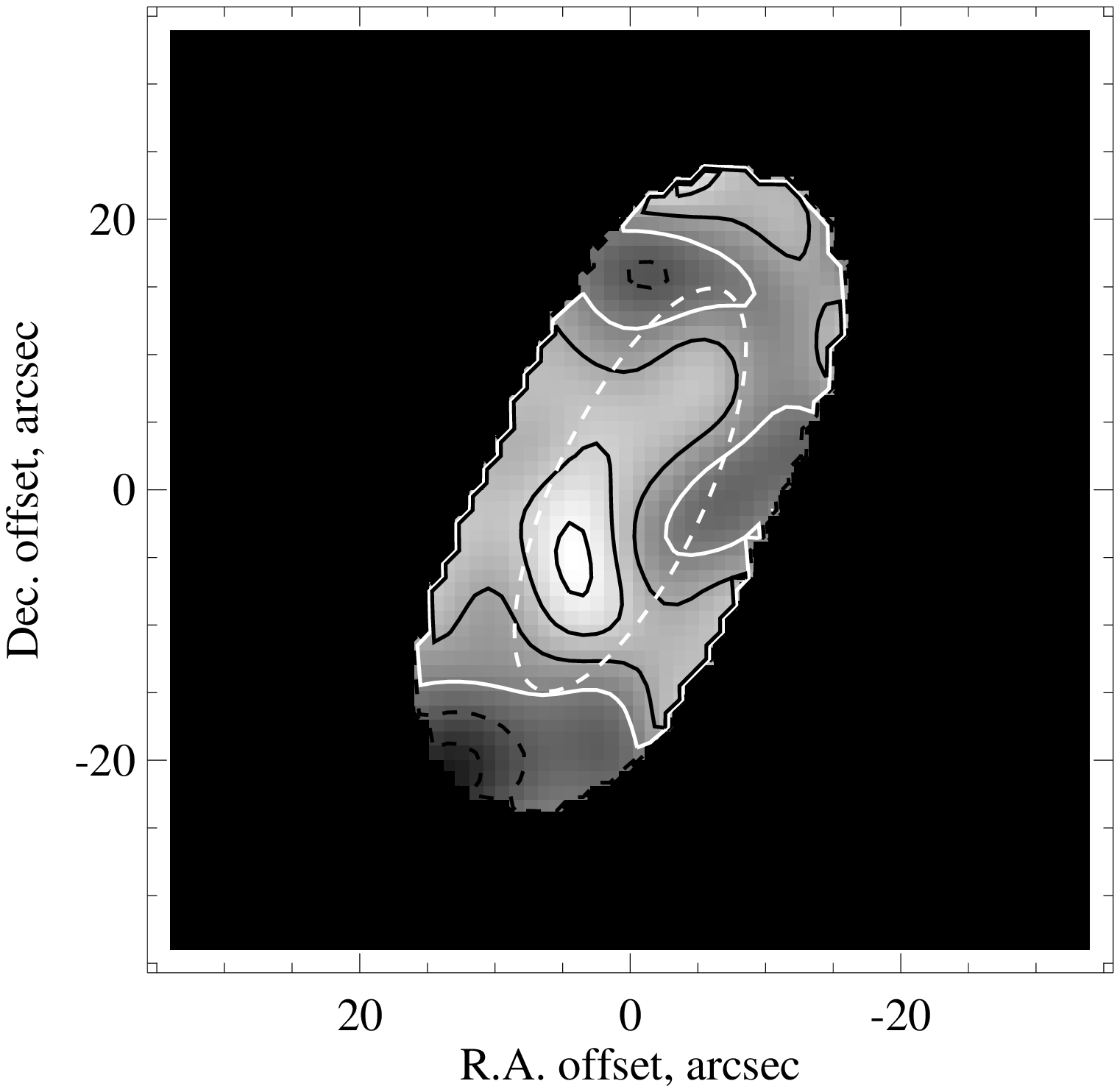,height=2.2in}
    \end{tabular}
    \caption{450 $\mu$m images of the Fomalhaut disk (adapted from
    Holland et al.~2002):
    (a) observation taken with SCUBA at the JCMT,
    (b) axisymmetric smooth disk model, and
    (c) the residuals (= observation - model), which show that the
    asymmetry in the observation could be explained by a bright clump
    embedded in the smooth disk.
    All contours are spaced at $1\sigma=13$ mJy beam$^{-1}$ levels.
    For (a) and (b) the lowest (solid white) contour is set at $3\sigma$,
    while for (c) the solid white line indicates the zero level and the
    black solid and dashed lines are the positive and negative $1\sigma$
    levels respectively.
    The dashed white oval shows the approximate inner edge
    of the mid-plane of the disk, a 125 AU radius ring inclined at
    $20^\circ$ to the line of sight.
    The coordinates are given as their offsets from the position of
    the star.
    The stellar photosphere has been subtracted from the observation.}
    \label{fig:fomim}
  \end{center}
\end{figure}

The photometric observations of emission from the Fomalhaut disk
are summarized in Table \ref{tab:fomsed}, and this SED is plotted in
Fig.~\ref{fig:fomsed}.
Recently the structure of the disk was mapped at both 450 and 850 $\mu$m
using SCUBA at the JCMT (Holland et al.~2002; Fig.~\ref{fig:fomim}a).
The double-lobed emission feature seen in the images implies that the disk
is being observed close to edge-on and that the dust is constrained to a
narrow ring $\sim 150$ AU from the star.
This is consistent with the disk's SED, which implies emission from cool
dust at a single temperature, as well as with previous observations
of the disk's structure (Holland et al.~1998; Harvey \& Jeffrys 2000).
The images at both wavelengths also show an asymmetry in the brightness
distribution.
The spatial structure of the disk seen in the 450 $\mu$m image
(Fig.~\ref{fig:fomim}a) was modelled in Holland et al.~(2002), where they
showed that the observation is consistent with emission from a smooth
axisymmetric ring (Fig.~\ref{fig:fomim}b) embedded within which is a bright
clump with a flux equal to $\sim 5$ per cent of the total disk flux
($\sim 30$ mJy; Fig.~\ref{fig:fomim}c).
Holland et al.~(2002) argue that this clump is real and associated with the
star, as the asymmetry is significantly above the noise at both wavelengths,
and there is a low probability of contamination from a background
source.

In the modelling presented here, we use the model presented in Holland et
al.~(2002) for the smooth ring.
This has a radial distribution of cross-sectional area $\bar{\sigma}(r)$
-- where $\bar{\sigma}(r)$d$r$ is the fraction of the total cross-sectional area
$\sigma_\rmn{tot}$ between $r$ and $r+$d$r$ --
that is indicative of a ring that has a mean radial distance of $\sim 150$ AU
and a width of $\sim 50$ AU.
This model is however described by the distribution of the orbital
elements of the material in the ring (see, e.g., WDT99).
Material in their model has average orbital inclinations and
eccentricities of $I=5^\circ$ and $e=0.065$.
While this modelling could not uniquely constrain
these parameters, these values result in a best fit to the data
and are consistent with indications later in this paper
that we are observing a collisional cascade for which such high values
are likely (section 5.3).

\begin{table}
  \begin{center}
    \caption{Photometric observations of the excess emission from Fomalhaut, $F_\nu$,
    and the $1\sigma$ errors, d$F_\nu$.
    The photospheric contribution has been subtracted from the observed
    emission using a Kurucz model atmosphere scaled to $K=0.992$ (Bouchet, Manfroid \&
    Schmider 1991).
    The \textit{IRAS} fluxes were taken from the Faint Source Catalog and a colour
    correction applied based on the temperature of the excess at each wavelength.
    This temperature is not well known at 25 $\mu$m and the resulting
    uncertainty in the derived flux is included in the errors.
    The \textit{ISO} fluxes were taken from Walker et al.~(in preparation).}
    \begin{tabular}{cccl}
       \hline
       $\lambda$, $\mu$m & $F_\nu$, Jy  & d$F_\nu$, Jy  & Reference \\
       25  & 1.24  & 0.6  & \textit{IRAS} \\
       60  & 9.80  & 0.62   & \textit{IRAS} \\
       60  & 7.51  & 0.92  & \textit{ISO} \\
       80  & 8.86  & 0.87  & \textit{ISO} \\
       100 & 10.95 & 0.66   & \textit{IRAS} \\
       100 & 11.04 & 2.0   & \textit{ISO} \\
       120 & 9.77  & 0.98  & \textit{ISO} \\
       150 & 7.12  & 0.67  & \textit{ISO} \\
       170 & 8.18  & 0.64  & \textit{ISO} \\
       200 & 3.55  & 1.15  & \textit{ISO} \\
       450 & 0.595 & 0.035 & Holland et al.~(2002) \\
       850 & 0.097 & 0.005 & Holland et al.~(2002) \\
       \hline
    \end{tabular}
    \label{tab:fomsed}
  \end{center}
\end{table}

\section{Theoretical Planetesimal Size Distribution}

\subsection{Basic idea}
Debris disks are thought to have formed in the first 10 Myr or so of
a star's life through canonical planet formation processes, wherein
collisions between initially sub-$\mu$m dust particles resulted in their
coagulation and eventual growth into planetesimals and maybe even planets
(e.g., Lissauer 1993).
Toward the end of this process the collisional processing was reversed
whereupon collisions between planetesimals more frequently resulted in
their net destruction rather than growth.
Such a change is likely to have been the consequence of an increase in
the average relative velocity of collisions, probably caused by some
combination of the removal of the gas disk (which was acting to damp the
eccentricities and inclinations of disk material) and gravitational
perturbations from planets or large planetesimals that formed in the
disk (Kenyon \& Bromley 2001), although this cross-over stage is not
well understood.
In the debris disk systems, we know that by the end of this process,
regardless of whether planets or planetesimals formed close
to the star, a disk of planetesimals was able to form and subsequently
survive in a ring at $>30$ AU from the star (Lagrange et al.~2001).
Once their collisional velocities are pumped up, these planetesimals
collide and break up into smaller fragments, initiating a collisional
cascade.
A typical planetesimal in such a cascade is most likely to have been
created by the break-up of a larger parent body in a collision with
another large planetesimal.
This parent body would most likely have been created by the break-up of
an even larger body, and the planetesimal itself will most likely end up
as a parent body for planetesimals smaller than itself.
It is the large quantities of small dust in the cascade, and its
commensurate large surface area, which makes a debris disk observable.
This dust is replenished as long as there remains a supply of large
primordial bodies to feed the cascade.

\subsection{Size distribution of an infinite collisional cascade}
The equilibrium size distribution resulting from a collisional cascade
can be found from theoretical arguments (Dohnanyi 1969):
\begin{equation}
  n(D) \propto D^{2-3q_\rmn{d}},
  \label{eq:nd}
\end{equation} 
where $n(D)$d$D$ is the number of planetesimals of size between $D$ and
$D+$d$D$.
For a self-similar collisional cascade it can be shown that $q_\rmn{d}=1.833$
(Tanaka, Inaba \& Nakazawa 1996), however if the strength of a planetesimal
is size dependent (see section 5), then the slope of the equilibrium
distribution could be slightly shifted from 1.833 (see fig.~12 of Durda
\& Dermott 1997; hereafter DD97).

\subsection{Size distribution of small grains}
The collisional cascade distribution is only expected to hold for disk
particles that are large enough not to be affected by radiation
forces (WDT99).
Particles smaller than a few $\mu$m are blown out of the
system by radiation pressure on orbital timescales and all
particles spiral into the star due to Poynting-Robertson (P-R) drag
on timescales that are proportional to their size.
It can be shown, however, that the high densities of the observed
extrasolar disks result in collisional lifetimes for dust of all sizes
that are shorter than their P-R drag lifetimes and so the latter
effect can be ignored (WDT99).
Thus here we make the assumption that the size distribution holds
down to the radiation pressure blow-out limit $D_\rmn{min}$
(i.e., those for which $\beta = F_\rmn{rad}/F_\rmn{grav} > 0.5$),
below which the distribution is cut-off.

A sharp cut-off below $D_\rmn{min}$ is, however, a simplification.
For a start, a sharp cut-off would cause a "wave" in the equilibrium size
distribution (Durda, Greenberg \& Jedicke 1998; hereafter DGJ98), since in
an equilibrium situation particles close to the blow-out limit should have
been broken up by the missing population of particles smaller than
themselves.
Since these are not removed their number increases and this affects
the particles they are expected to break up, and so on.
Also, while it may be possible to ignore the contribution of particles
smaller than the blow-out limit to the total cross-sectional area of disk
material (e.g., if their lifetime is much shorter than their collisional
lifetime, WDT99), the size distribution of gravitationally bound grains
close to the blow-out limit could be affected by their destruction in
collisions with the blow-out grains (Krivov, Mann \& Krivova 2000).
The size distribution of these bound grains would be further complicated
by radiation pressure which causes them to have large eccentricities,
thus reducing their number density in the disk region (see, e.g.,
fig.~4 of Augereau et al.~2001).

\subsection{Size distribution of large planetesimals}
At the large size end, planetesimals that have a collisional lifetime
that is longer than the age of the system do not form part of the
collisional cascade, and their distribution must be primordial.
However it is unclear what this would be, as even the primordial
distribution of solar system objects is unknown.
It is uncertain whether the distribution of asteroids larger than
30 km is primordial due to uncertainties in the collisional properties
of these asteroids (DGJ98).
The distribution of Kuiper belt objects larger than 100 km,
which appear to have a relatively steep distribution ($q_\rmn{d}=1.83$--2.27,
Luu \& Jewitt 1998; Gladman et al.~1998), is also difficult to interpret,
since while the objects themselves might be primordial, their distribution
may not be.
The sweeping of Neptune's resonances through the young Kuiper belt as
Neptune's orbit expanded early in the history of the solar
system (Hahn \& Malhotra 1999) would have resulted in a size
distribution of objects trapped in resonance that is steeper
than the primordial distribution (Jewitt 1999).
The primordial distribution expected from theoretical grounds is
also unclear, since simulations of planet formation result in
distributions that are usually steeper than the cascade distribution,
but with $q_\rmn{d}$ anywhere between 2 and 2.5 (e.g., Kokubo \& Ida 2000;
Kenyon 2002).
In any case, the size distribution of the largest planetesimals in the
distribution is likely to be dominated by the runaway growth process,
whereby these grew faster than the rest of the population, causing a
bimodal size distribution (e.g., Wetherill \& Stewart 1993; Kokubo \&
Ida 1996).
Runaway growth is discussed further in section 7.2.

\section{SED Modelling}

\subsection{The SED model}
A disk's SED is determined by three factors:
the spatial distribution of material in the disk,
its size distribution,
and its optical properties (determined by its composition).
While these factors could be interrelated, we assume here
that they are independent.
For the spatial distribution we assume that particles of all
sizes have the spatial distribution inferred from the modelling of
the 450 $\mu$m images defined by $\bar{\sigma}(r)$ (see section 2).
For the size distribution we consider one that follows
equation (\ref{eq:nd}), and assume that this distribution is
cut-off below $D_\rmn{min}$, the radiation pressure blow-out limit,
and extends up to large enough sizes that this limit does not
affect the SED.

For the particles' composition we use the core-mantle model developed
by Li \& Greenberg (1997) for interstellar dust, which has since been
applied to cometary dust (Greenberg 1998) and to dust grains in
extrasolar disks (Li \& Greenberg 1998; Augereau et al.~1999).
It seems reasonable to apply this model to extrasolar disks, since these
grains are likely to be reprocessed (albeit significantly) interstellar
grains.
In this model the dust is assumed to be aggregates of core-mantle grains
with a silicate core ($\rho = 3500$ kg m$^{-3}$) and an organic
refractory mantle comprised of UV photo-processed ices which accreted
onto the silicate cores in the interstellar medium ($\rho = 1800$ kg
m$^{-3}$).
These aggregates are assumed to be porous and to have a fraction of
water ice ($\rho = 1200$ kg m$^{-3}$) filling the gaps;
this is ice that would have frozen onto the growing grains in the
protoplanetary disk.
Since interstellar grains are amorphous and would only be
crystallized at the high temperatures expected close to stars,
we assume that both the silicate and ice components of the grains
are amorphous.
We also assume a volume fraction of silicates to organic refractory
material in the core-mantle of 1:2 as found for interstellar-type
grains (Li \& Greenberg 1997).
Thus the particles' composition is defined by their porosity, $p$,
and the volume fraction of the gap in the grains that is filled with
ice, $q_\rmn{ice}$.
We use the optical constants for each of the components taken from
Li \& Greenberg (1997, 1998) and compute those of the composite
material using Maxwell-Garnett Effective Medium theory (Bohren
\& Huffman 1983).
The absorption efficiencies of these grains, $Q_\rmn{abs}(\lambda,D)$,
are then calculated using Mie theory (Bohren \& Huffman 1983),
Rayleigh-Gans theory or geometric optics in the appropriate
limits (see Laor \& Draine 1993).
These then define the temperature of the grains, $T(D,r)$,
and their radiation pressure blow-out limit, $D_\rmn{min}$
(see, e.g., WDT99).

Thus the SED model is defined by the following four parameters:
$\sigma_\rmn{tot}$ (the total cross-sectional area in the disk),
$q_\rmn{d}$, $p$, and $q_\rmn{ice}$.
The flux in Jy from the disk at a given wavelength is given by (e.g., WDT99)
\begin{eqnarray}
  F_\nu & = & 2.35\times10^{-11}R_\star^{-2}\sigma_\rmn{tot}
    \int_{D_\rmn{min}}^{D_\rmn{max}} Q_\rmn{abs}(\lambda,D)\bar{\sigma}(D)
    \nonumber \\
        &   & \times \int_{r_\rmn{min}}^{r_\rmn{max}} B_\nu[T(D,r)]
              \bar{\sigma}(r)\rmn{d}r \rmn{d}D,
  \label{eq:fnu}
\end{eqnarray}
where $\bar{\sigma}(D) = 0.25\pi D^2n(D)/\sigma_\rmn{tot}$
and $\sigma_\rmn{tot}$ is in AU$^2$.
Here we have assumed that the disk is optically thin at all wavelengths
and that there is no disk self-heating.

\subsection{The modelling procedure}
We determined the best fit $q_\rmn{d}$ for each particle composition
(defined by $p$ and $q_\rmn{ice}$) by minimizing
$\chi^2=\sum_{i=1}^{N_\rmn{obs}}
[(F_{\nu_i, \rmn{obs}}-F_{\nu_i, \rmn{mod}})/dF_{\nu_i, \rmn{obs}}]^2$,
where $\sigma_\rmn{tot}$ was set to scale the SED to fit the 850 $\mu$m
flux.
Fig.~\ref{fig:fomsed} shows how this procedure was applied for   
solid grains (for which $p=0$, $D_\rmn{min} = 7$ $\mu$m,
$\rho = 2370$ kg m$^{-3}$, and which by definition contain no water ice
from the protoplanetary disk).
The best fit gave $q_\rmn{d} = 1.84$, $\sigma_\rmn{tot} = 33.7$ AU$^2$,
and $\chi^2 = 43.4$.
This type of modelling constrains $q_\rmn{d}$ very well since it has a large
effect on the SED:
increasing $q_\rmn{d}$ increases the contribution of small grains and so
results in an increase in the flux shortward of about 60 $\mu$m
(due to the higher temperatures of small grains) but a decrease in the
flux at longer wavelengths (since small grains emit less efficiently than
larger ones at long wavelengths)\footnote{It is not immediately obvious
from Fig.~\ref{fig:fomsed} that this cross-over occurs at 60 $\mu$m
because, as these models have been scaled to $F_\nu(850$ $\mu$m),
those with a higher $q_\rmn{d}$ also have a higher $\sigma_\rmn{tot}$.}.

Increasing $q_\rmn{d}$ also steepens the sub-mm spectral slope, defined by
$\alpha$, where $F_\nu \propto \lambda^{-\alpha}$, because
the emission here is dominated by grains of a size comparable with
the wavelength so that the ratio of fluxes at two wavelengths is
determined by the ratio of cross-sectional area in grains of size
comparable to those wavelengths.
This effect can be quantified by a simple model in which all the dust
is at the same distance from the star and has the same temperature, $T$,
and has emission efficiencies that are given by
\begin{equation}
  Q_\rmn{abs}(\lambda,D) = \left\{ \begin{array}{ll}
      1             & \mbox{for $\lambda<D$} \\
      (\lambda/D)^n & \mbox{for $\lambda>D$}
      \end{array} \right.
  \label{eq:sqmod}
\end{equation}
Putting this into equation (\ref{eq:fnu}), we find that in the Rayleigh-Jeans
regime:
\begin{equation}
  F_\nu \propto \left\{ \begin{array}{ll}
      \lambda^{3-3q_\rmn{d}}
      & \mbox{for $n>3q_\rmn{d}-5$} \\
      \lambda^{-2-n}
      & \mbox{for $n<3q_\rmn{d}-5$};
      \end{array} \right.
  \label{eq:sfnumod}
\end{equation}
i.e., increasing $q_\rmn{d}$ steepens the slope up to a constant value for
$q_\rmn{d} > 5/3 + n$.
Thus the sub-mm spectral slope is largely determined by the size distribution
of particles of sub-mm sizes.
This demonstrates the importance of an SED model not only giving a good
$\chi^2$ fit to the whole SED, but also giving a good fit to
$\alpha_\rmn{obs} = 2.85$.
In this simple model, the observed slope implies that either $q_\rmn{d}=1.95$
and $n > 0.85$, or $n=0.85$ and $q_\rmn{d}>1.95$.
In fact we find that for solid grains $\alpha -\alpha_\rmn{obs} \approx
4.9(q_\rmn{d} - 1.86)$ and so $q_\rmn{d} = 1.84$ fits the sub-mm slope
relatively well.
The discrepancy from the simple model is because real grains have emission
features that are not reproduced by equation (\ref{eq:sqmod}).

\subsection{Modelling results}

\begin{figure}
  \begin{center}
    \begin{tabular}{c}
      \psfig{figure=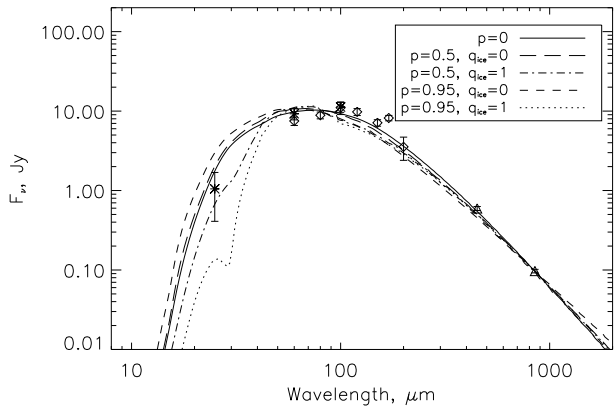,height=2.1in}
    \end{tabular}
    \caption{The best fits to the observed SED for five models
    with different grain compositions (see legend).
    The resulting best fits for the size distribution, the total
    cross-sectional area in the disk, and the $\chi^2$ for the
    five models (top to bottom in the legend) are:
    $q_\rmn{d} = $ [1.84,1.81,1.87,1.81,1.88],
    $\sigma_\rmn{tot} = $ [33.7,33.2,27.1,34.7,23.4] AU$^2$,
    and $\chi^2 = $ [43.4,75.3,101,118,117].
    The minimum grain size in these models is:
    $D_\rmn{min} = $ [7,13,9,124,11] $\mu$m.}
    \label{fig:fomsedmods}
  \end{center}
\end{figure}

The analysis in Fig.~\ref{fig:fomsed} was repeated using grains
with a range of porosities and ice fractions, but found that
increasing either $p$ or $q_\rmn{ice}$ resulted in a worse fit
to the data (i.e., a higher $\chi^2$, see
Fig.~\ref{fig:fomsedmods}).
The problem with increasing the porosity of the grains is that
even though solid grains are less affected by radiation pressure
than are porous grains (mostly due to their higher density), and
so can exist in the disk down to much smaller sizes (see
$D_\rmn{min}$ in the caption of Fig.~\ref{fig:fomsedmods}),
the disk's smallest porous grains are still hotter than its
smallest solid grains because they appear to be made up of an
agglomeration of much smaller particles.
Thus the SED resulting from a porous-grained disk peaks at shorter
wavelengths than a solid-grained one (see Fig.~\ref{fig:fomsedmods}).
Higher porosity grains also have fewer emission features which
results in a flatter sub-mm slope.
The problem with increasing the ice fraction in the grains is that
they have a range of emission features in the far-IR, but not of
a shape that increases the sub-mm slope to the same levels as for non-icy
grains.
The resulting spectrum is too peaked in the 40--90 $\mu$m wavelength region
(see Fig.~\ref{fig:fomsedmods}), and because icy grains are inefficient
absorbers of starlight, their cool temperature results in a poor fit to
the mid-IR flux.
A poor fit to the 25 $\mu$m flux would not in itself rule out a model,
since there may be a contribution to the observed flux from an, as yet
undetected, hotter disk component such as that inferred to exist in the
HR 4796 system (Augereau et al.~1999).
We conclude that while solid non-icy grains are the most likely
candidates for the grains in this disk, SED modelling cannot set
strict constraints on the composition of the grains without a more
accurate determination of the disk's mid-IR emission.
However, a more robust feature of the SED modelling is the size
distribution parameter $q_d$, which lies in the range 1.81--1.88
for all models.

\begin{figure}
  \begin{center}
    \begin{tabular}{c}
      \psfig{figure=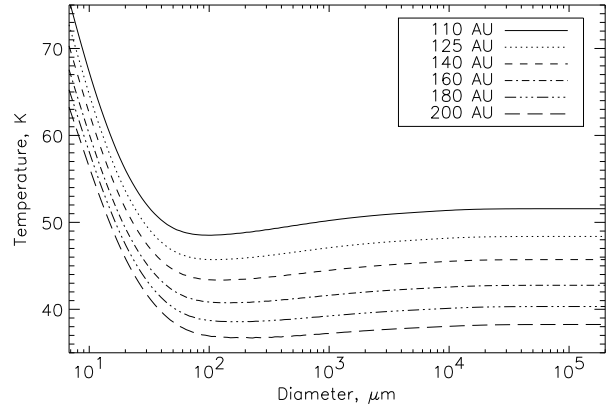,height=2.1in}
    \end{tabular}
    \caption{The temperatures of different sized dust grains at
    different distances from Fomalhaut, assuming these are solid
    grains composed of 1/3 (by volume) amorphous silicate and 2/3
    organic refractory material.}
    \label{fig:fomtemps}
  \end{center}
\end{figure}

In the rest of this paper we use the parameters ($D_\rmn{min}$,
$\rho$, $q_\rmn{d}$ and $\sigma_\rmn{tot}$) derived from this
modelling assuming solid non-icy grains, since this composition
is also consistent with that expected on theoretical grounds.
Grains that grow through grain--grain collisions in a protoplanetary
disk are expected to have significant porosity (e.g., Wurm \& Blum
1998), similar to that of grains resulting from the sublimation
of comets in the solar system ($p=0.85$--0.95).
Grain--grain collisions, if energetic enough however, can lead to
significant grain compaction (Dominik \& Tielens 1997).
Thus we expect grains produced by comet sublimation to have a
high, primordial-like, porosity (such as has been inferred for the
grains around around $\beta$ Pictoris and HR 4796A; Li \& Greenberg
1998, Augereau et al.~1999), whereas grains at the end of a
collisional cascade should be more compact, maybe with a porosity
close to that of stone meteorites ($p = 0.05$--0.3; Flynn, Moore \&
Kl\"{o}ck 1999).
The size distribution we have inferred, $q_\rmn{d}$ close to 1.84,
is consistent with the dust having its origin in a collisional cascade
(section 3.2);
collisional lifetime arguments in the following section also support
this hypothesis.
Also, we expect little contribution to the dust population
from cometary sublimation because of the low temperatures in
the disk (see Fig.~\ref{fig:fomtemps}).
Thus a low level of porosity is to be anticipated for the dust.

It is harder to predict the amount of water ice on the grains.
Certainly the low temperatures in the disk imply that while the grains
were forming (i.e., before they were compacted by collisions), water
vapour should have condensed onto the aggregates, at least partially
filling the gaps in the grains (Pollack et al.~1994; Li \& Greenberg
1998).
However, this ice may have since been sublimated by repeated
collisional heating, or removed by sputtering by stellar wind ions
(Jurac, Johnson \& Donn 1998).
Or indeed this ice may still exist on the grains, yet it could look
more like the organic refractory material in our model due to
the formation of an irradiation mantle by stellar UV photons and
galactic cosmic rays (e.g., Johnson et al.~1987; Cooper, Christian
\& Johnson 1998; Jewitt 1999).
We note that water ice features have been detected in some, but
not all, Kuiper belt objects (e.g., Luu, Jewitt \& Trujillo 2000;
Brown, Blake \& Kessler 2000).

\subsection{Limitations for large planetesimals}

\begin{figure}
  \begin{center}
    \begin{tabular}{c}
      \psfig{figure=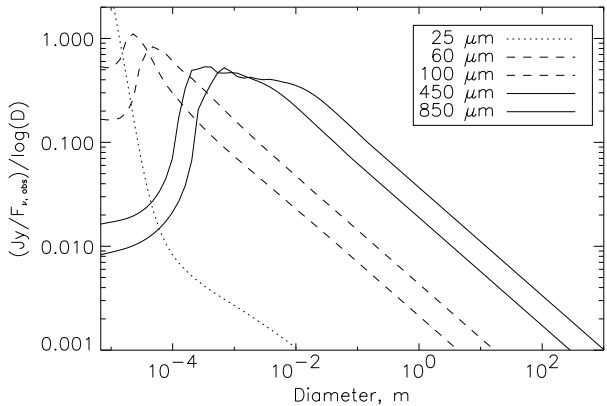,height=2.1in}
    \end{tabular}
    \caption{The contribution of different sized dust grains to the observed
    fluxes for the best fit model to the SED (i.e., that shown in
    Fig.~\ref{fig:fomsed} with $q_\rmn{d} = 1.84$).
    The area under the curves between two grain sizes defines the
    proportion of the flux that originates from this size range;
    the total area across all sizes is unity.}
    \label{fig:psd}
  \end{center}
\end{figure}

The size distribution inferred from the SED modelling can only be applied
to grains that contribute to the observed SED fluxes.
Fig.~\ref{fig:psd} shows that planetesimals larger than $\sim 0.2$ m
in diameter contribute less than 5 per cent of the 850 $\mu$m flux
(and even less at shorter wavelengths).
Thus we can only derive the size distribution for grains that are
smaller than $\sim 0.2$ m (although, as we will see in the next section,
we can extend this up to larger sizes by collisional lifetime arguments).
Since the total mass of the Fomalhaut disk, assuming that the size distribution
with $q_\rmn{d}=1.84$ extends up to $D_\rmn{max}$, is given by
\begin{equation}
  m_\rmn{tot}/M_\oplus \approx 0.5D_\rmn{max}^{0.48}, \label{eq:mtot}
\end{equation}
where $D_\rmn{max}$ is in m, this means that we see a dust mass of
$\sim 0.2 M_\oplus$.

\section{Catastrophic Collisions}
The outcome for a planetesimal of size $D$ when it is impacted by
another of size $D_\rmn{im}$ is determined by the specific incident
kinetic energy,
\begin{equation}
  Q = 0.5(D_\rmn{im}/D)^3v_\rmn{col}^2. \label{eq:q}
\end{equation}
The shattering threshold, $Q_\rmn{S}^\star$, is defined as the specific
incident energy required to break up the planetesimal, resulting in fragments
with a range of sizes up to half the mass of the original planetesimal.
Collisions with $Q<Q_\rmn{S}^\star$ result in cratering whereby some material
is ejected, but the planetesimal remains largely intact.
More energetic collisions result in the planetesimal being reduced
to more and smaller fragments\footnote{For collisions with very small relative velocities
there are two other possible collisional outcomes: the planetesimals could
rebound in an inelastic collision, or they could stick together.
While these possibilities are important in protostellar disks, they are not
considered in this paper, since the collisional velocities in debris disks
are thought to be too high for these modes to be important.}.
In general larger planetesimals have lower shattering thresholds, since
they contain larger flaws which are activated at lower tensile stresses
(Housen \& Holsapple 1990, 1999).
Very large planetesimals, however, are strengthened by their gravitational
self-compression which inhibits crack propagation (Davis et al.~1985).
After the collision, the fragments disperse due to the kinetic energy
imparted to them in the collision.
For the break-up of a large planetesimal ($D>150$ m), however, this energy
may not be enough to overcome the gravitational binding
energy and some fragments may reaccumulate into a rubble pile
(e.g., Campo Bagatin, Petit \& Farinella 2001; Michel et al.~2001).
A catastrophic collision is defined as one in which $Q>Q_\rmn{D}^\star$,
where the dispersal threshold, $Q_\rmn{D}^\star$, refers to an impact in which
the largest fragment resulting from the collision after reaccumulation has
taken place (i.e., this fragment could be a rubble pile) has half the mass
of the original planetesimal.
The break-up of a small planetesimal (i.e., $D<150$ m), for which
$Q_\rmn{D}^\star \approx Q_\rmn{S}^\star$, is said to occur in the
strength regime, while the break-up of a larger planetesimal is said
to occur in the gravity regime, where $Q_\rmn{D}^\star \gg Q_\rmn{S}^\star$.

A planetesimal's catastrophic collisional lifetime depends on the disk's
size distribution, since only collisions in which $D_\rmn{im} > D_\rmn{tc}
= X_\rmn{tc}D$ have enough energy to be destructive, where
\begin{equation}
  X_\rmn{tc} = (2Q_\rmn{D}^\star/v_\rmn{tc}^2)^{1/3}. \label{eq:xcc}
\end{equation}
We use the term \textit{threshold--catastrophic} to denote a collision
between planetesimals of size $D$ and $D_\rmn{tc}(D)$, and $v_\rmn{tc}(D)$ is
the relative velocity of such collisions.
Here we assume that all collisions have the same average relative velocity
at large separations\footnote{The distributions of the eccentricities and
inclinations of planetesimals predicted from planet formation models are
dependent on their size.
However, in a debris disk undergoing a collisional cascade these
distributions would depend on the mechanism which causes the high
eccentricities and inclinations in these disks;
e.g., if this mechanism is the continual gravitational stirring of
massive planetesimals or planets in the disk, these distributions
would be uniform with planetesimal size, whereas if this mechanism was
stirring by a passing star these distributions would evolve with
time after the event and would be size dependent
(Kenyon \& Bromley 2002).},
\begin{equation}
  v_\rmn{rel} = f(e,I)v_\rmn{k}, \label{eq:vrel}
\end{equation}
where $f(e,I)$ is a function of the average eccentricities and
inclinations of the planetesimals given as $\sqrt{1.25e^2 +I^2}$
(Lissauer \& Stewart 1993; Wetherill \& Stewart 1993), and
$v_\rmn{k}$ is the keplerian velocity at this distance from the star.
In the Fomalhaut model $f(e,I) \approx 0.11$;
thus $v_\rmn{rel} \approx 0.4$ km s$^{-1}$ at the mean distance of 150 AU.
The relative velocity of a collision is then
\begin{equation}
  v_\rmn{col}^2 = v_\rmn{rel}^2+ v_\rmn{esc}^2(D,D_\rmn{im}), \label{eq:vim}
\end{equation}
where the mutual escape velocity of the two planetesimals is given by
\begin{equation}
  v_\rmn{esc}^2(D,D_\rmn{im}) = (2/3)\pi G\rho
    \frac{D^3+D_\rmn{im}^3}{D+D_\rmn{im}}. \label{eq:vesc}
\end{equation}
The increase in impact velocity due to gravity (i.e., gravitational focussing)
becomes important for collisions for which $v_\rmn{esc} > v_\rmn{rel}$, i.e.,
for planetesimals with $D>v_\rmn{rel}/\sqrt{(2/3)\pi G\rho}$.
For Fomalhaut this corresponds to planetesimals larger than about 700 km
in diameter, or 0.6 per cent of a lunar mass.

\subsection{Dispersal threshold model}
Several authors have studied how $Q_\rmn{S}^\star$ and $Q_\rmn{D}^\star$ vary
with planetesimal diameter for a variety of materials (e.g., rocks/ice)
with a range of structural properties (e.g., homogeneous/rubble pile),
and for collisions with a range of relative velocities (e.g.,
km s$^{-1}$ debris disk-like/m s$^{-1}$ protostellar nebula
disk-like impacts) and impact parameters (e.g., head-on/glancing blow).
The techniques that have been used to study collisions range from
lab experiments which provide a direct measure of a collisional
outcome (e.g., Fujiwara et al.~1989; Housen \& Holsapple 1999),
to theoretical studies (e.g., Petit \& Farinella 1993),
to interpretation of the distributions of the asteroid families (e.g., Tanga
et al.~1999; Cellino et al.~1999) or the main-belt population (DD97; DGJ98),
to computational modelling using smooth particle hydrodynamics (SPH; Love
\& Ahrens 1996; Benz \& Asphaug 1999, hereafter BA99).
However, most of these studies are valid only in specific regimes
(e.g., in the strength or gravity regimes).

In this paper we use the results of BA99 who use SPH to model the
fragmentation of both solid ice and basalt of a broad range of sizes (up to
200 km diameter) for impacts with a range of impact parameters and
with relative velocities comparable to those expected in extrasolar
planetesimal disks.
In the rest of this paper $Q_\rmn{D}^\star$ refers to the dispersal
threshold averaged over all impact parameters (BA99).
The BA99 models reproduce the well-known features of the $Q_\rmn{D}^\star$
vs $D$ plot;
i.e., the dispersal threshold decreases with size ($\propto D^{-(0.36-0.45)}$)
in the strength regime due to the lower shattering strength of larger
planetesimals, but increases with size ($\propto D^{1.19-1.36}$)
in the gravity regime due to the extra energy required to impart
enough kinetic energy to the collisional fragments to overcome the
planetesimal's gravity.
However, the impact strengths they derive for cm-sized ice grains
impacted at $\sim 3$ km s$^{-1}$ are 20--50 times higher than those
directly measured in the laboratory (e.g., Arakawa 1999; Ryan, Davis \&
Giblin 1999).
Their results for basalt impacted at 3 km s$^{-1}$ do not show the
same discrepancy (Holsapple 1994).

\begin{figure}
  \vspace*{0.0in}
  \begin{center}
    \begin{tabular}{rr}
       \textbf{(a)} & \psfig{figure=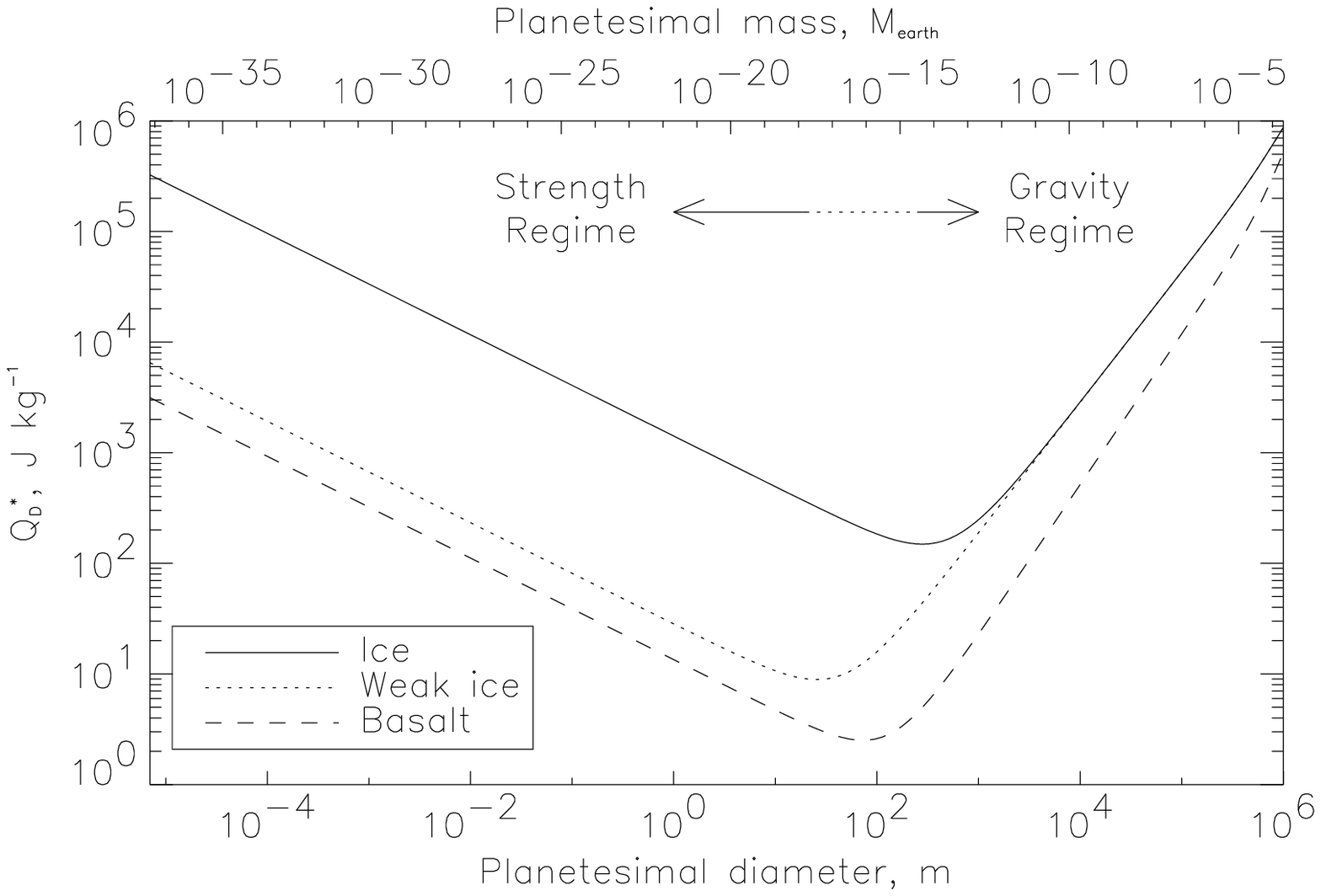,height=1.88in} \\[0.15in]
       \textbf{(b)} & \psfig{figure=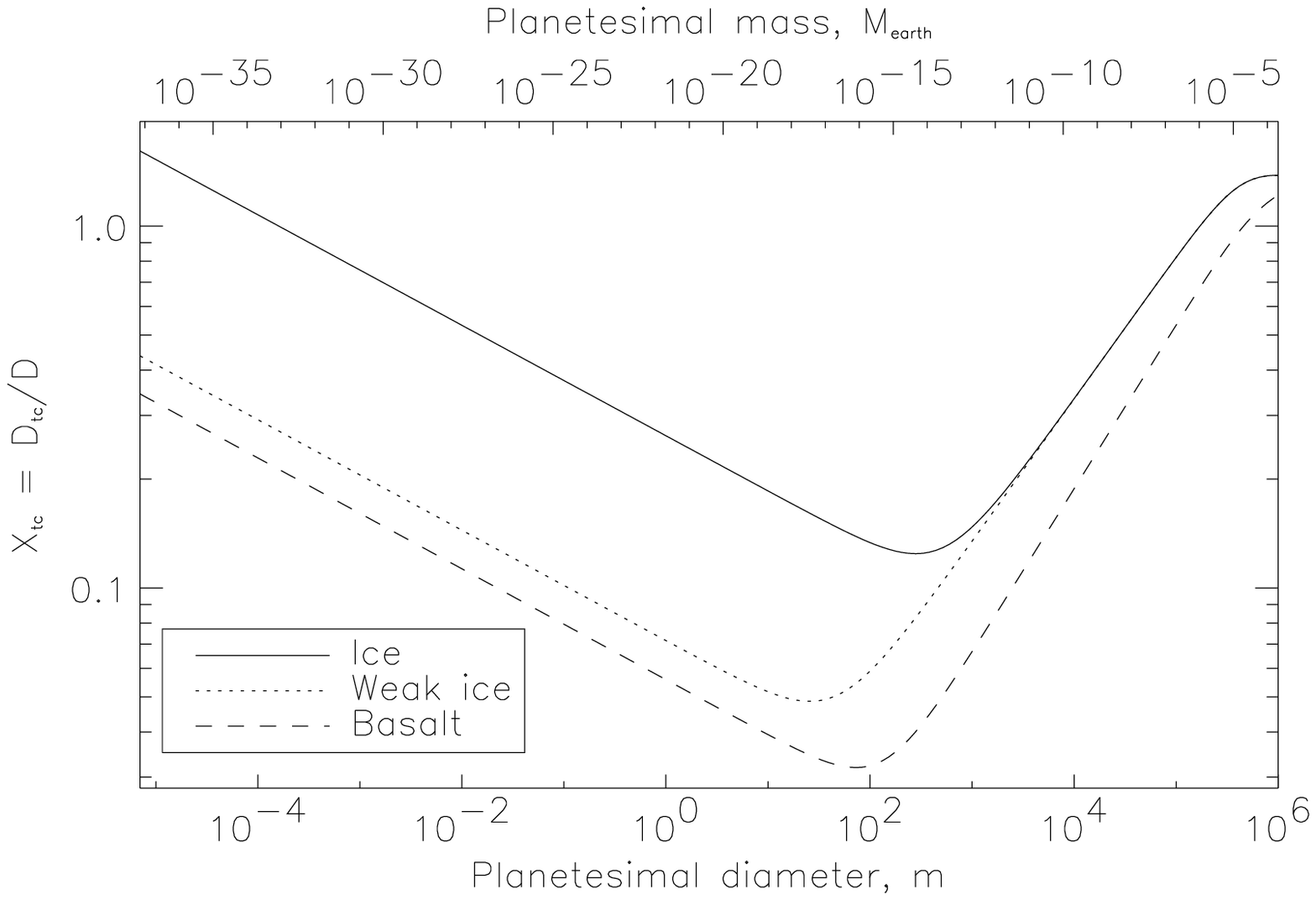,height=1.88in}
    \end{tabular}
    \caption{The collisional properties of planetesimals of different
    sizes in the Fomalhaut disk for the three models (Ice, Weak Ice, and
    Basalt) based on the SPH modelling of Benz \& Asphaug (1999):
    (a) the dispersal threshold;
    (b) the relative size of an impactor required to catastrophically
    break up a planetesimal.
    The dispersal threshold decreases with size in the strength  
    regime ($D<150$ m) due to the lower shattering strength of larger
    planetesimals, but increases with size in the gravity regime due to
    the extra energy required to impart enough kinetic energy to the
    collisional fragments to overcome the planetesimal's gravity.
    The relative size required for a catastrophic impact varies with $D$
    as does $Q_\rmn{D}^\star$ except that it flattens off for $D>700$ km
    because the extra energy required to disperse the collisional fragments
    is provided by the higher impact velocities, which have been enhanced
    by the planetesimals' gravitational attraction (from $\sim 0.4$ km s$^{-1}$
    at 100 km to $\sim 0.8$ km s$^{-1}$ at 1000 km).}
    \label{fig:collprops}
  \end{center}
\end{figure}

While the SED modelling suggested that there is little water ice in the
Fomalhaut grains, the structural properties of the organic refractory
mantle would be close to that of the ice studied in BA99, since the mantle
is essentially comprised of UV photoprocessed ices.
Thus the collisional properties of planetesimals in Fomalhaut's disk would
lie somewhere between those of rock and ice.
Here we consider three different models for $Q_\rmn{D}^\star$:
\textbf{Ice} refers to the BA99 result for ice impacted at 0.5 and
3 km s$^{-1}$;
\textbf{Weak Ice} also refers to the BA99 results for ice, except that
the part of a planetesimal's $Q_\rmn{D}^\star$ resulting from its strength
is reduced by a factor of 50 to reflect the discrepancy with laboratory
results;
and \textbf{Basalt} refers to the BA99 result for basalt impacted at
3 and 5 km s$^{-1}$.
In all cases we have used $\rho = 2370$ kg m$^{-3}$ and the BA99 results
at the two impact velocities have been scaled to the relative velocity
of a threshold-catastrophic collision, $v_\rmn{tc}$,
assuming that $Q_\rmn{D}^\star \propto v_\rmn{col}^{\gamma(D)}$.
This procedure is iterative since to calculate $Q_\rmn{D}^\star$ we need
to know $v_\rmn{tc}$, for which we need to know $X_\rmn{tc}$, which is
itself determined by both $Q_\rmn{D}^\star$ and $v_\rmn{tc}$.
The resulting models for $Q_\rmn{D}^\star$ and $X_\rmn{tc}$ 
are shown in Fig.~\ref{fig:collprops}.

While the BA99 models apply only to planetesimals up to 200 km in
diameter, we have assumed that their results can also be scaled up to
1000 km diameter.
We do not expect their models to be applicable much beyond this,
since in this regime collisions occur at the planetesimals' escape
velocity.
As some energy is lost in the collision, only a small fraction of
the collisional fragments can escape, and so such collisions must
result in net accretion rather than destruction
(i.e., $Q_\rmn{D}^\star \rightarrow \infty$).
Such an effect would not have been picked up in BA99, since their
experiments were all carried out with impact velocities that are
more than an order of magnitude under the escape velocities.
Collisions between such massive bodies, which further we expect to be
rubble piles, are discussed in section 7.

\subsection{Catastrophic collision timescale}
The rate of impacts from planetesimals in the size range $D_\rmn{im}$ to
$D_\rmn{im}+$ d$D_\rmn{im}$ falling onto a planetesimal of size $D$ is given
by $R_\rmn{col}(D,D_\rmn{im})$d$D_\rmn{im}$, where (e.g., Opik 1951)
\begin{equation}
  R_\rmn{col}(D,D_\rmn{im}) = f(D,D_\rmn{im})\sigma(r,\theta,\phi)v_\rmn{rel}.
  \label{eq:rcol}
\end{equation}
In this expression, $\sigma(r,\theta,\phi)$ is the average cross-sectional
area density of disk material, which, for a ring of radius $r$, radial width
d$r$, and that has a (small) mean orbital inclination of particles in the ring
of $I$, is approximately
\begin{equation}
  \sigma(r,\theta,\phi) = \sigma_\rmn{tot}/(4\pi r^2\rmn{d}rI). \label{eq:sigrtp}
\end{equation}
Also, $f(D,D_\rmn{im})\sigma(r,\theta,\phi)$d$D_\rmn{im}$ is the cross-sectional
area density of planetesimals in the given size range that the planetesimal of size
$D$ sees, where
\begin{equation}
  f(D,D_\rmn{im}) = \bar{\sigma}(D_\rmn{im})
    \left( \frac{D+D_\rmn{im}}{D_\rmn{im}} \right)^2
    \left[ 1+ \frac{v_\rmn{esc}^2(D,D_\rmn{im})}{v_\rmn{rel}^2} \right].
    \label{eq:fddim}
\end{equation}
The last term in equation (\ref{eq:fddim}) is the gravitational focussing factor,
which becomes important for collisions with planetesimals in the Fomalhaut
disk that are larger than about 700 km.

A planetesimal's catastrophic collision timescale is thus given by
\begin{equation}
  t_\rmn{cc}(D) = t_\rmn{per}(rdr/\sigma_\rmn{tot})[2I/f(e,I)]/f_\rmn{cc}(D),
  \label{eq:tcc}
\end{equation}
where $t_\rmn{per}$ is the orbital period at $r$, and
$f_\rmn{cc}(D)$ is defined by
\begin{equation}
  f_\rmn{cc}(D) = \int_{D_\rmn{tc}(D)}^{D_\rmn{max}} f(D,D_\rmn{im})
    \rmn{d}D_\rmn{im}, \label{eq:fcc}
\end{equation}
where $D_\rmn{tc}$ is the larger of $X_\rmn{tc}D$ or $D_\rmn{min}$.

\begin{figure}
  \begin{center}
    \begin{tabular}{c}
       \psfig{figure=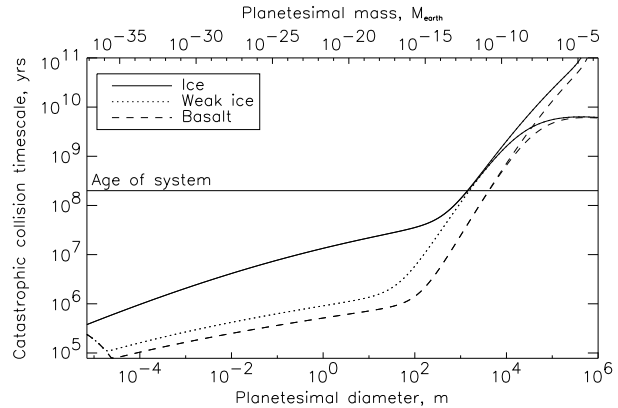,height=2.1in}
    \end{tabular}
    \caption{The collisional lifetime of planetesimals of different
    sizes in the Fomalhaut disk.
    The solid, dotted and dashed lines are for the three models for the collisional
    properties of these planetesimals described in the text.
    The two lines for each model assume that the size distribution inferred
    from the SED modelling ($q_\rmn{d} = 1.84$) extends up to planetesimals of
    size 1000 and 10,000 km
    (distributions with larger planetesimals have lower collisional lifetimes).}
    \label{fig:tccs}
  \end{center}
\end{figure}

Fig.~\ref{fig:tccs} shows the catastrophic collisional lifetime of
planetesimals of different sizes in the Fomalhaut disk for the
three models of their collisional properties, using the
size distribution inferred from the SED modelling ($q_\rmn{d} = 1.84$)
assuming that this is truncated at 1000 and 10,000 km.
Normally a planetesimal is broken up by one that just has enough energy to
do so (i.e., in a threshold-catastrophic collision), since of all the
planetesimals that can destroy it, these have the largest total
cross-sectional area in the disk.
However, the enhanced cross-sectional area of gravitationally focussing
planetesimals can mean that a planetesimal is more likely to end its life
by being accreted onto one of these than by being destroyed by a smaller
planetesimal, hence the shorter collisional lifetime of large planetesimals
when $D_\rmn{max} = 10,000$ km.
As there is less cross-sectional area available to destroy larger
planetesimals, there is a steady increase in collisional lifetime with size;
e.g., a simple model which assumes that most of this area is in
planetesimals of size $X_\rmn{tc}D$, where $X_\rmn{tc} < 1$, shows that if
$Q_\rmn{D}^\star \propto D^a$ then $t_\rmn{cc} \propto
D^{(3+a)q_\rmn{d}-(5+a)}$, implying that the BA99 models should have
$t_\rmn{cc} \propto D^{0.14-0.22}$ in the strength regime and
$t_\rmn{cc} \propto D^{1.52-1.66}$ in the gravity regime, consistent with
Fig.~\ref{fig:tccs}.

\subsection{Implications for Fomalhaut's size distribution}
While the collisional lifetime of all planetesimals contributing to
the Fomalhaut observations ($D < 0.2$ m) is model dependent, because each
requires a collision with different sized planetesimals for catastrophic
destruction (Fig.~\ref{fig:collprops}b), all models predict a lifetime
that is shorter than the age of the system (200 Myr) by more than an order
of magnitude (Fig.~\ref{fig:tccs}).
Thus the primordial populations of this size range of particles has already
been substantially depleted.
However, these populations will also have been replenished with collisional
fragments from the break-up of larger planetesimals.
Since the largest particles that we can see are replenishing the smaller
particles, \textit{it is an unavoidable conclusion that this part of the size
distribution forms a collisional cascade}.
The real questions are:
\textbf{(i)} is the cascade in equilibrium, and
\textbf{(ii)} how far does it extend up in size, or in other words,
what is feeding the cascade?

In answer to \textbf{(i)}, while the parameterization of the size distribution
assumed in the SED modelling is only an approximation (e.g., sections 3.2 and
3.3), the derived slope,$q_\rmn(d) = 1.84$, is very close to that expected
for an equilibrium cascade.
Also, the short collisional lifetimes in Fig.~\ref{fig:tccs} imply that
the cascade should have reached equilibrium within the age of the
system.

In answer to \textbf{(ii)}, the short collisional lifetime of the
0.2 m diameter planetesimals we see today (0.4--9 Myr depending on
their composition, Fig.~\ref{fig:tccs}) implies that if there were no
larger planetesimals with which to replenish their number, then while
all 0.2 m planetesimals we see must be primordial, their current
population should only be a small fraction of their original
population.
In fact, it must be an infinitely small fraction of the original population,
since the collisional lifetime of these planetesimals would have been
even shorter at earlier epochs when their population was more massive
(this lifetime is approximately inversely proportional to its mass).
In other words, unless the collisional cascade was only initiated
in the last few Myr, the large number of 0.2 m planetesimals we observe
today implies that there must be a population of larger bodies.
Similar arguments can be applied for all planetesimals for which
their collisional lifetime is shorter than the time since collisions
became destructive (which we assume here is approximately the age of
the system);
the frequency of collisions amongst these large bodies means that
their distribution would also be expected to follow that
of a collisional cascade.
Thus we infer that the collisional cascade starts with planetesimals
of a size for which their collisional lifetime is equal to the age of
the system, i.e., those with $D_\rmn{max(cc)}=1.5$--4 km
(Fig.~\ref{fig:tccs}), implying that the mass of material in the
collisional cascade is $m_\rmn{cc} \approx 20$--30 $M_\oplus$
(equation \ref{eq:mtot}).
If we had assumed that collisions became destructive more recently than
200 Myr ago, we would have inferred the cascade to start with smaller
planetesimals.
However, we note that a few km is also the size of planetesimal thought
to have fed the collisional cascade in the young ($<1$ Gyr) Kuiper
belt (Kenyon 2002).

We also predict that any planetesimals larger than 1.5--4 km are
predominantly primordial and so their size distribution is
also primordial;
these will have contributed little to the collisional cascade.
Their size distribution can be somewhat constrained from considerations
of the total mass of the disk, which is dominated by that of the
largest bodies.
If the primordial size distribution connects smoothly to the collisional
cascade distribution, and can be defined by the power law exponent
$q_\rmn{p}$ between $D_\rmn{max(cc)}$ and $D_\rmn{max}$, then the
total mass of primordial material would be
\begin{equation}
  m_\rmn{p}/m_\rmn{cc} = 0.48(6-3q_\rmn{p})^{-1}[(D_\rmn{max}/
    D_\rmn{max(cc)})^{6-3q_\rmn{p}}-1]. \label{eq:mp}
\end{equation}
The minimum mass solar nebula (Weidenschilling 1977), if it
extended out to such large distances, would have had a mass of solid
material of 0.1--0.2 $M_J$ in the region 125--175 AU.
Planet formation models predict that for the gravitational perturbations
of large planetesimals in the disk to have stirred the disk sufficiently
to ignite a collisional cascade within 200 Myr, the planetesimals in
Fomalhaut's disk would have had to have grown to at least 1000 km in
diameter and the disk must be at least 10 times more massive than the
minimum mass solar nebula (see fig.~5 of Kenyon \& Bromley 2001).
Thus a reasonable distribution for the primordial planetesimals, and one
which we will use for reference later in the paper, is one with
$q_\rmn{p} = 1.84$ and $D_\rmn{max} = 1000$ km, since
such a disk would have a mass of $\sim 1.1$ Jupiters, i.e., 5--10
times that of the minimum mass solar nebula.
We note that 1000 km is similar in size both to the largest asteroids
and Kuiper belt objects (Trujillo et al.~2001) as well as to the
runaway growth limit of $\sim 700$ km derived from the 
proposed eccentricities and inclinations of the orbits of
Fomalhaut's planetesimals (e.g., section 5).
The distribution of runaway planetesimals is discussed in section 7.

\begin{figure}
  \begin{center}
    \begin{tabular}{c}
       \psfig{figure=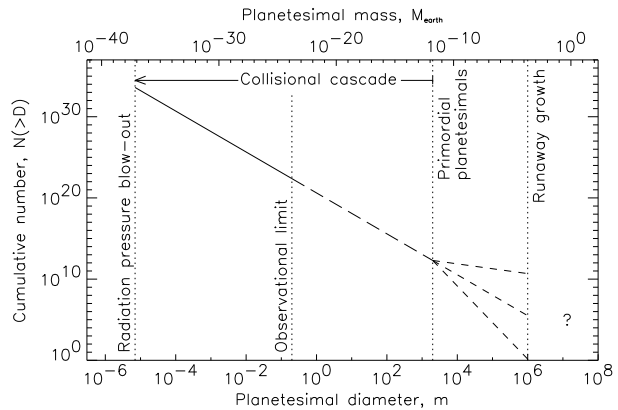,height=2.1in}
    \end{tabular}
    \caption{Inferred size distribution of planetesimals in Fomalhaut's
    disk:
    A collisional cascade extends from $\sim 4$ km planetesimals down
    to 7 $\mu$m dust grains.
    Smaller dust grains are blown out of the system by radiation
    pressure.
    At sub-mm and shorter wavelengths, we only see those members smaller
    than 0.2 m.
    Planetesimals larger than $\sim 4$ km have an unknown primordial size
    distribution, however a reasonable distribution may be one that
    extends up to 1000 km with the same slope as the cascade (see
    section 5.3).
    Planetesimals larger than $\sim 1000$ km, if they
    exist, must have grown by runaway growth and would form a separate
    population (see section 7.2).}
    \label{fig:sizedistrn}
  \end{center}
\end{figure}

We note that the shape of the primordial size distribution determines
how the brightness of the disk evolves with time, and that this
evolution can be constrained using the distribution of the
brightnesses of observed debris disks.
However we leave this discussion to a future paper.
Our understanding of the size distribution of planetesimals in
Fomalhaut's disk from this study is summarized in
Fig.~\ref{fig:sizedistrn}.

\section{Intrinsic Clumpiness}
Collisions between planetesimals cause the disk to be intrinsically
clumpy, since after a collision the fragments of the target and the impactor
move away from the impact site at a velocity determined by the kinetic energy
imparted to them in the collision.
This means that the debris from each planetesimal forms an expanding clump the
centre of which follows the orbit of the parent planetesimal.
At any one time we would expect to see clumps in the disk with a range
of both fluxes (those from different magnitude collisions) and sizes
(smaller ones that were created recently and older, more extended, clumps).
The probability of our witnessing a collisional clump of a given brightness
and physical size depends on:
the rate of collisions between different sized planetesimals (section 6.1);
the amount of dust such collisions produce, which depends on the size
distribution of collisional fragments (section 6.2);
and on the rate at which this dust precesses around the orbit of the parent
planetesimal, which depends on the ejection velocity of collisional fragments
(section 6.3).

\subsection{Collision rates}
The number of collisional events per unit time that occur in the disk
between (target) planetesimals of size $D$ to $D+$d$D$ and those (impactors)
of size between $D_\rmn{im}$ and $D_\rmn{im}+$d$D_\rmn{im}$ is
$R_\rmn{col}(D,D_\rmn{im})$d$D_\rmn{im}n(D)$d$D$, where
the rate of collisions onto individual planetesimals, $R_\rmn{col}$,
was defined in equation (\ref{eq:rcol}).
These collision rates are most strongly affected by the disk's size
distribution and the total amount of cross-sectional area in it,
but are largely unaffected by the eccentricities and inclinations of
disk material, as $R_\rmn{col} \propto \sqrt{1+1.25e^2/I^2}$.
These orbital parameters do, however, have a strong effect on the
outcome of these collisions.

\subsection{Collisional clump flux}
The size distribution of collisional fragments was not discussed in BA99,
since it was not permitted by the resolution of their models.
Here we use the results of Campo Bagatin \& Petit (2001; hereafter CP01)
which considers the size distribution of collisional fragments expected
from geometrical constraints.
In the CP01 model, a planetesimal of mass $M$ is broken up sequentially into
smaller fragments, where the size of each fragment is determined by
the shape of the volume remaining after the previously created fragments
have been removed.
The physical basis of this approach is that fragmentation proceeds by the
coalescence of flaws that propagate through the target after impact.
The size distribution they find depends on the mass of the largest remnant
created in the collision, $f_\rmn{lr} = M_\rmn{lr}/M$.
However, apart from the few largest fragments, the distribution was found to
follow a power law with an index $q_\rmn{c} \approx 1.91-1.97$, in agreement
with their analytical model which predicted a distribution with $q_\rmn{c}=1.93$.
This size distribution also concurs with experiments which find that while
$q_\rmn{c}$ can fall anywhere in the range 1.6--2.6 (Fujiwara et al.~1989; Giblin et
al.~1998; Housen \& Holsapple 1999), values close to 1.9--2 are more common.
While it is not clear why geometrical arguments should apply in the gravity
dominated regime, since here each fragment is a gravitationally
reaccumulated conglomerate of the smaller fragments created in the
collision, CP01 proved that their model reproduces the size distributions
of the asteroid families, the members of which should be reaccumulated
fragments (Campo Bagatin et al.~2001), and which have observed slopes
of $q_\rmn{c} = 1.83$--2.17.

Here we assume that the size distribution of fragments smaller than the
second largest remnant, that with diameter $D_2$, follows a power law
size distribution with $q_\rmn{c}=1.93$ down to a lower limit $D_\rmn{min, cl}$.
It may be inappropriate to extrapolate the CP01 size distribution to small
fragments, since their distribution may be dominated by different
physics to that of larger fragments.
For example, Durda \& Flynn (1999) found a knee in the size distribution of
dust created by the break-up of composite materials due to flaws at material
boundaries.
This knee occurs at the size of crystals embedded in the planetesimal,
($\sim 1$ mm in size in their experiment).
A flattening of the size distribution for dust smaller than 1 mm was
also reported by Fujiwara et al.~(1989).
In any case, we note that while fragments smaller than the radiation
pressure blow-out limit, $D_\rmn{min}$, may be created in collisions,
they would be removed from the disk within an orbital timescale
by radiation pressure.
Thus we set $D_\rmn{min, cl} > D_\rmn{min}$.

Conserving the volume of the original planetesimal, and scaling the
fragment distribution such that there are two fragments larger than $D_2$,
we find that
\begin{equation}
  D_2/D = \left[ \left( \frac{2-q_\rmn{c}}{q_\rmn{c}} \right)
          (1-f_\rmn{lr}) \right]^{1/3}. \label{eq:d2d}
\end{equation}
The total amount of cross-sectional area created by the break-up
of a planetesimal is thus
\begin{eqnarray}
  \sigma_\rmn{col}(D,D_\rmn{im}) & = & \sigma_\rmn{min}C_{q_\rmn{c}}
    [1-f_\rmn{lr}(D,D_\rmn{im})]^{q_\rmn{c}-1} \nonumber \\
    & & \times (D/D_\rmn{min, cl})^{3q_\rmn{c}-3},
  \label{eq:scoll}
\end{eqnarray}
where $C_{q_\rmn{c}} = [(2q_\rmn{c}-2)/(q_\rmn{c}-5/3)](2/q_\rmn{c}-1)
^{q_\rmn{c}-1} = 0.323$.
Equation (\ref{eq:scoll}) is only valid when $D_2<D_\rmn{lr}$, i.e.,
when $f_\rmn{lr} > 1-0.5q_\rmn{c} = 0.035$.
When $f_\rmn{lr} \leq 0.035$ we assume that there is more than one
fragment of size $D_\rmn{lr}$, and that the $q_\rmn{c}=1.93$ distribution
applies for smaller fragments.
Thus we get a new expression describing the outcome of such a
super-catastrophic collision:
\begin{eqnarray}
  \sigma_\rmn{col}(D,D_\rmn{im}) & = & \sigma_\rmn{min}[(2-q_\rmn{c})/(q_\rmn{c}-5/3)]
    f_\rmn{lr}(D,D_\rmn{im})^{q_\rmn{c}-2} \nonumber \\
    & & \times (D/D_\rmn{min, cl})^{3q_\rmn{c}-3}.
  \label{eq:scoll2}
\end{eqnarray}
Note that the cross-sectional area given in equations (\ref{eq:scoll}) and
(\ref{eq:scoll2}) describe that of the debris produced by one of the
planetesimals involved in the impact, that which we call the target,
and that in both cases this cross-sectional area is
$\propto D_\rmn{min, cl}^{-0.79}$.

BA99 found that the size of the largest remnant of a collision, whether rubble
pile or intact fragment, is determined by the collisional energy in the
following way: $f_\rmn{lr} = 0.5-s(Q/Q_\rmn{D}^\star-1)$,
where $s$ is close to 0.5 for all materials and impact velocities,
and $Q_\rmn{D}^\star$ is the dispersal threshold at the relevant collisional
velocity (i.e., not averaged over all impact parameters as before).
This is consistent with the results of experiments of
catastrophic impacts in the strength regime, which found that
$f_\rmn{lr} = 0.5(Q_\rmn{D}^\star/Q)^{1.24}$ (Fujiwara et al.~1989).
It is also consistent with the results of experiments of
cratering impacts in the strength regime:
e.g., Holsapple (1993) used scaling arguments to show that in the strength
regime, $M_\rmn{crater}/M = 1 - f_\rmn{lr} \propto
(Q/v_\rmn{col}^2)(v_\rmn{col}^2/Q_\rmn{S}^\star)^{1.5\mu}$;
for $\mu = 2/3$ and scaling to $f_\rmn{lr} = 0.5$ when $Q=Q_\rmn{S}^\star$,
we find that $f_{lr} = 1-0.5Q/Q_\rmn{S}^\star$.
For collisions in which $f_\rmn{lr} > 0.5$, we use the BA99 results
with $s=0.5$ so that
\begin{equation}
  f_\rmn{lr}(D,D_\rmn{im}) = 1-0.5 \left[ \frac{D_\rmn{im}}{X_\rmn{tc}(D)D} \right]^3
                     \left[ \frac{v_\rmn{col}(D,D_\rmn{im})}{v_\rmn{tc}(D)}
                     \right]^{2-\gamma(D)}, \label{eq:flr2}
\end{equation}
where we remind the reader that $Q_\rmn{D}^\star \propto v_\rmn{col}^{\gamma(D)}$.
For more energetic collisions, however, we use the experimental results summarized
in Fujiwara et al.~(1989), since these cover a wider range of collisional outcomes,
and BA99 only considered collisions in which $f_\rmn{lr} \approx 0.5$.
Thus for $f_\rmn{lr} < 0.5$, we use
\begin{equation}
  f_\rmn{lr}(D,D_\rmn{im}) =
    0.5 \left[ \frac{X_\rmn{tc}(D)D}{D_\rmn{im}}                \right]^{3.72}
        \left[ \frac{v_\rmn{tc}(D)} {v_\rmn{col}(D,D_\rmn{im})} \right]^{2.48-1.24\gamma(D)},
    \label{eq:flr3}
\end{equation}
which is consistent with the BA99 results in this regime.
Here we assume that $f_\rmn{lr}$ averaged over all impact parameters
can be obtained using the averaged $Q_\rmn{D}^\star$ model of section 5.
We also set the physical constraint that $f_\rmn{lr} \geq (D_\rmn{min}/D)^3$.

\begin{figure}
  \begin{center}
    \begin{tabular}{c}
       \psfig{figure=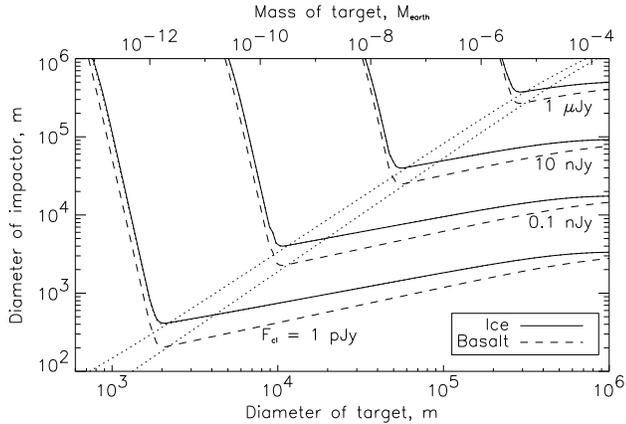,height=2.2in}
    \end{tabular}
    \caption{The size of impactor planetesimals in the Fomalhaut
    disk required to break up the target into fragments that emit a given
    flux, $F_\rmn{cl}$, at 450 $\mu$m.
    The fragments are assumed to follow a power law with $q_\rmn{c} = 1.93$
    down to the radiation pressure blow-out limit of 7 $\mu$m.
    The solid and dashed lines correspond to the models for the
    collisional properties of \textbf{Ice} and \textbf{Basalt} described in
    the text;
    in this target size range \textbf{Weak Ice} has the same properties as
    \textbf{Ice}.
    The four lines for each model correspond to $F_\rmn{cl} =$ 1 $\mu$Jy, 10 nJy,
    0.1 nJy and 1 pJy.
    The dotted lines demark the limits of threshold catastrophic collisions
    for the \textbf{Ice} and \textbf{Basalt} models.
    Collisions below these lines result in cratering from the target planetesimal,
    while those above result in the complete disintegration of the target.}
    \label{fig:fomccls}
  \end{center}
\end{figure}


Consider the dust clumps created in collisions between planetesimals
in the Fomalhaut disk.
In the following discussion, the emission from these clumps is calculated
from equation (\ref{eq:fnu}) assuming that the composition of this dust
is the same as that inferred for the smooth ring from the SED
modelling of section 4, and that this material is spread across the same
range of distances from the star as the smooth ring (i.e., 150 $\pm 25$ AU).
Fig.~\ref{fig:fomccls} shows the size of impactor required to produce
debris clumps of target material that emit a range of clump fluxes at 450
$\mu$m, assuming a fragment cut-off at $D_\rmn{min, cl} = D_\rmn{min}$.
This shows that for cratering events (i.e., for collisions below
the dotted lines), the amount of dust produced depends mostly
on the size of the impactor, not on the size of the target;
i.e., whether an impactor falls on a planetesimal much larger
than itself or one closer to its own size, a similar amount of
dust is released from the target.
Similarly, the amount of dust resulting from the catastrophic
destruction of a target planetesimal (in a collision above the
dotted lines) is largely independent of the size of
planetesimal impacting the target.
It is also interesting to note that debris from the smaller
planetesimal in a cratering collision can be brighter than
that from the larger planetesimal.

If we had chosen the size distribution to be cut-off at
$D_\rmn{min, cl} = 1$ mm, instead of 7 $\mu$m, then while the cross-sectional
area resulting from a given collision would be decreased by a factor of
$(1000/7)^{-0.79}$, the flux per unit cross-sectional area would have
increased, because while the extra 7 $\mu$m to 1 mm particles contain
a lot of cross-sectional area, they don't emit efficiently at 450 $\mu$m.
As the clump's 450 $\mu$m flux in Jy is $6.5 \times 10^{-3}$ times
its cross-sectional area in AU$^2$ if the cut-off is at 7 $\mu$m,
but is 0.11 times this area if the cut-off is at 1 mm, this means that
with this larger cut-off, the fluxes in Fig.~\ref{fig:fomccls} should
be reduced by a factor of $\sim 3$.

\begin{figure}
  \begin{center}
    \begin{tabular}{c}
       \psfig{figure=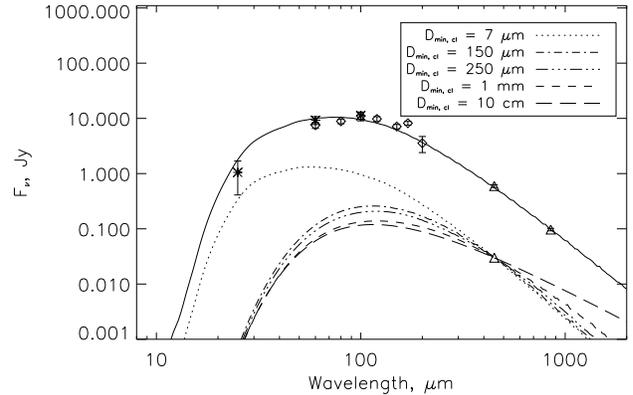,height=2.05in}
    \end{tabular}
    \caption{The SED of emission from a clump in Fomalhaut's disk
    assuming it is composed of grains with a size distribution with
    $q_\rmn{c} = 1.93$ that is cut-off at the small size end
    below $D_\rmn{min, cl}$.
    The clump's SED is normalised to 30 mJy at 450 $\mu$m, which
    is the flux of the clump observed by Holland et al.~(2002; see
    Fig.~\ref{fig:fomim}c).
    Also shown is both the observed SED (symbols) and
    the fit to this SED assuming all this emission comes from
    the smooth ring (solid line).}
    \label{fig:fomsedcldmin}
  \end{center}
\end{figure}

To determine what the clumps in Fig.~\ref{fig:fomccls}, which are
defined by their 450 $\mu$m emission, would emit at other
wavelengths, consider Fig.~\ref{fig:fomsedcldmin}, which shows the
SED of their emission normalised to 30 mJy at 450 $\mu$m (the flux
of the observed clump).
This SED is shown for different $D_\rmn{min, cl}$ as its shape
is strongly dependent on this cut-off due to the increased temperature
of small grains (Fig.~\ref{fig:fomtemps}).
For example, collisions corresponding to 1 $\mu$Jy at 450 $\mu$m for a clump
with a cut-off at 7 $\mu$m (i.e., those close to the top right of
Fig.~\ref{fig:fomccls}) would result in emission of 11 $\mu$Jy at 25 $\mu$m
and 43 $\mu$Jy at 60 $\mu$m, while if fragments from the same collisions
had a cut-off at $D_\rmn{min, cl} = 1$ mm, this would result in emission
of 0.3 $\mu$Jy at 450 $\mu$m, just 7 nJy at 25 $\mu$m, and 0.7 $\mu$Jy at
60 $\mu$m.
Note that if the collisional size distribution extends down to the
blow-out limit, the clump would appear brighter relative to the smooth
ring at mid- to far-IR wavelengths than in the sub-mm.

\subsection{Clump growth rates}
The rate at which the physical extent of a clump grows is determined by
the ejection velocity of collisional fragments.
This is usually discussed in terms of the parameter
$f_\rmn{KE}$, the fraction of the impact energy, $E_\rmn{col} =
0.5MM_\rmn{im}v_\rmn{col}^2/(M+M_\rmn{im})$, that ends up as kinetic
energy of the fragments, a parameter that is not well known.
Laboratory experiments of impacts of cm-sized objects imply that the
kinetic energy imparted to the largest fragments produced in
the collision (i.e., those containing 70--80 per cent of the target mass)
is a fraction $f_\rmn{KE} = 0.3$--3 per cent of the impact energy
(Fujiwara \& Tsukamoto 1980), with an unknown quantity imparted to
smaller fragments.
Studies of the asteroid families, on the other hand, imply that
family-forming events had $f_\rmn{KE} \approx 0.1$ (e.g., Davis et
al.~1989).
However, SPH simulations of impacts in the gravity regime imply
even higher values of $f_\rmn{KE} = 0.2$--0.4 (Love \& Ahrens 1996).
It has also been reported that $f_\rmn{KE}$ increases as the velocity
of the impact increases $\propto Q^{0.5}$ (e.g., Fujiwara \&
Tsukatmoto 1980; Paolicchi et al.~1996), and it
is expected that collisions with rubble pile planetesimals
have a lower $f_\rmn{KE}$ than those with
solid bodies due to the inefficient coupling of impact energy
(e.g., Campo Bagatin et al.~2001).
While clearly a simplification, in this paper we assume that
$f_\rmn{KE} = 0.1$, and that this is true in both the strength and
gravity regimes.
We also assume that this energy is split evenly between the target
and the impactor, since this was found to be the case both for
cratering events in the strength regime (Hartmann 1988) and for
collisions in the gravity regime (Love \& Ahrens 1996), although
we recognize that the division of this energy depends on the impact
parameter (Love \& Ahrens 1996) among other things.

The kinetic energy, $f_\rmn{KE}E_\rmn{col}$,
is not evenly distributed across the fragments.
Fragments are found to attain a range of velocities, the distribution of
which can be approximated by the relation $f(v) \propto v^{-k}$, where
$f(v)$d$v$ is the fraction of material with velocities between $v$ and
$v+$d$v$.
The power law index, $k$, has been reported with values between
3.25 (Gault et al.~1963) and 1.5 (Love \& Ahrens 1996),
where values of $k<2$ imply that most of the kinetic energy is carried
away by a small mass of material which is moving very fast (e.g.,
Love \& Ahrens 1996; BA99) and which usually originates near the impact
site.
There also appears to be a dependence of the velocity of a fragment
on its mass in that, on average, lower mass fragments move faster
than more massive ones.
The mean velocity of fragments of size $D$ deduced from laboratory
experiments is reported to be $\propto D^{-r}$, where $r$ lies anywhere
between 0 (e.g, Giblin et al.~1998) and 0.5 (e.g., Nakamura \& Fujiwara
1991), although the scatter about this mean tends to dominate over any
strong trend (Giblin et al.~1998).
Very little of the kinetic energy is found to be imparted to
the largest remnant (Nakamura \& Fujiwara 1991; Michel et al.~2001).

Here we calculate the characteristic ejection velocity of target
fragments assuming that the available kinetic energy is distributed
among all fragments except the largest remnant;
i.e., $0.5f_\rmn{KE}E_\rmn{col} = 0.5(1-f_\rmn{lr})Mv_\rmn{ej}^2$,
so that
\begin{equation}
  v_\rmn{ej}^2(D,D_\rmn{im}) = \frac{f_\rmn{KE}Q(D,D_\rmn{im})/[1+(D_\rmn{im}/D)^3]}
                               {[1-f_\rmn{lr}(D,D_\rmn{im})]}.
  \label{eq:vej}
\end{equation}
This means that target fragments created in a cratering event, wherein
$f_\rmn{lr}$ is given by equation (\ref{eq:flr2}) and $D_\rmn{im} \ll D$, have
an ejection velocity of $v_\rmn{ej} \approx \sqrt{2f_\rmn{KE}Q_\rmn{D}^\star}$.
Thus cratering debris from a given target has the same ejection velocity
regardless of the size of the impactor.
This is because while the amount of kinetic energy imparted to fragments
is larger for larger impactors, there is also a greater mass of material
ejected from the largest remnant for this to be shared between.
This relation also means that the ejection velocity is lowest for
cratering events that occur at the transition between strength and gravity
scaling, since these are the weakest planetesimals (Fig.~\ref{fig:collprops}a).
Equation (\ref{eq:vej}) also means that the ejection velocity of the fragments
of the smaller planetesimal in the collision, for which $f_\rmn{lr} \approx 0$,
is $v_\rmn{ej} \approx \sqrt{0.5f_\rmn{KE}}v_\rmn{col}(D,D_\rmn{im})$;
i.e., the impactor retains a fixed fraction (22 per cent in this case)
of its original velocity.
This means that for planetesimals for which gravitational focussing is not
important, all impactor debris has the same ejection velocity,
which is $\sim 90$ m s$^{-1}$ in the Fomalhaut disk.

In the gravity regime, some of the kinetic energy imparted to the
fragments is converted into gravitational energy, since the fragments
have to overcome the gravity of the largest remnant.
The energy required to disperse all fragments except the largest remnant
is $E_\rmn{grav} = 1.2(GM^2/D)[1-f_\rmn{lr}^{5/3}]$.
Thus the characteristic velocity of the fragments once far from the largest
remnant, $v_\infty$, is given by
\begin{eqnarray}
  v_\infty       & = & \sqrt{v_\rmn{ej}^2-v_\rmn{grav}^2}, \\
  v_\rmn{grav}^2 & = & 0.4\pi G\rho D^2[1-f_\rmn{lr}(D,D_\rmn{im})^{5/3}]
                       \nonumber \\
                 &   & \times [1-f_\rmn{lr}(D,D_\rmn{im})]^{-1},
\end{eqnarray}
where this assumes that the kinetic energy given to the individual
fragments of the largest remnant is small.
However, using this model we find that for collisions between the
largest planetesimals ($D>100$ km) in the gravity scaling regime
in Fomalhaut's disk, $v_\rmn{ej}$ is sometimes slightly smaller
than $v_\rmn{grav}$.
This occurs when $Q_\rmn{D}^\star < (1/3) \pi \rho GD^2/f_\rmn{KE}$.
This discrepancy is due to our extrapolation of $Q_\rmn{D}^\star$
from the BA99 models to larger planetesimals and different impact
velocities.
We already know that the collisional model may not be applicable
to the break-up of planetesimals larger than $\sim 700$ km diameter
in this disk, and such collisions are discussed in section 7.
However, it may also be that $f_\rmn{KE}$ is higher in the gravity regime.
Setting $f_\rmn{KE} = 0.34$ would resolve the discrepancy is this size
range.
A higher $f_\rmn{KE}$ may be expected because a much smaller fraction of
the impact energy is used in shattering the target;
e.g., if $f_\rmn{KE} \propto (Q/Q_\rmn{S}^\star)^{0.5}$ (Fujiwara \&
Tsukamoto 1980), then threshold catastrophic collisions in the gravity
regime would have a very high $f_\rmn{KE}$ (although some of this
kinetic energy would go into the individual fragments of the largest
remnant before they reaccumulate, at which point it would be converted
into internal energy).
Here we account for this discrepancy by artificially increasing
$f_\rmn{KE}$ where required so that $v_\rmn{ej} \geq 1.01 v_\rmn{grav}$.
The reasoning behind this is that in the gravity regime the fragments
attain a range of velocities, and while those with
$v_\rmn{ej} < v_\rmn{grav}$ remain bound to the largest remnant, most
of those that escape must have $v_\rmn{ej}$ just larger than
$v_\rmn{grav}$.

The rate at which a clump appears to grow depends on the rate at which
the fragments precess around the orbit of the parent planetesimal.
The radial and vertical structure of the clump change with time as well,
however the spatial extent of these variations are at or below the level
of the azimuthal growth.
A fragment's precession rate is determined by the velocity with which it
is ejected in the direction of the orbital motion, d$v_\rmn{k}$, since
this causes a change in the semimajor axis of the fragment's orbit, and
hence in its orbital period, d$t_\rmn{per}/t_\rmn{per} =
-3$d$v_\rmn{k}/v_\rmn{k}$.
The ejection velocity field is generally found to be isotropic
both in experiments (Giblin et al.~1998) and analysis of the
orbital elements of the asteroid families (Zappala et al.~1996).
Thus the average velocity of all fragments in the direction of orbital
motion is d$v_\rmn{k} \approx v_\infty/\sqrt{3}$, and the rate of
growth of the total azimuthal extent of the clump is
\begin{equation}
  \dot{\theta} = 2\sqrt{3}v_\infty/a.
  \label{eq:thetadot}
\end{equation}

Before continuing we define the characteristic lifetime of a clump of target
fragments created in a collision between the target (of size $D$) and an
impactor (of size $D_\rmn{im}$), $t_\rmn{cl}(D,D_\rmn{im})$, to be equal
to the average time it takes for the clump fragments to occupy half the
ring (i.e., $\pi/\dot{\theta}$):
\begin{equation}
  t_\rmn{cl}(D,D_\rmn{im})/t_\rmn{per} = (1/4\sqrt{3})
    [v_\rmn{k}/v_\infty(D,D_\rmn{im})].
  \label{eq:tcl}
\end{equation}
Thus in this model impactor debris from collisions between planetesimals
in Fomalhaut's disk remains in a clump for about
6 orbital periods, while target debris clumps last for between 10 and 700
orbital periods, depending on the target's dispersal threshold
(clumps from weak, 10--1000 m, planetesimals last longest).

\subsection{Discussion}
Combining the collision rate from section 6.1 and the clump growth rate from
section 6.3, we find that the number of clumps we expect to see at any one
time that are smaller in azimuthal extent than $\theta$, and that were created
in collisions between (target) planetesimals of size $D$ to $D+$d$D$ and those
(impactors) of size between $D_\rmn{im}$ and $D_\rmn{im}+$d$D_\rmn{im}$ is
given by
\begin{equation}
  N(<\theta) = R_\rmn{col}(D,D_\rmn{im})\rmn{d}D_\rmn{im}n(D)\rmn{d}D
    (\theta/\dot{\theta}).
  \label{eq:nlth}
\end{equation}
We also know how bright those clumps should be, since this was
defined in section 6.2 (Fig.~\ref{fig:fomccls}).
Thus we can integrate over all impacts that create clumps brighter
than a specific limit to get the total number of clumps we expect to see
in the disk that are smaller in azimuthal extent than $\theta$ and brighter
than $F_\rmn{cl}$, $N(<\theta,>F_\rmn{cl})$.

Clumps that are greater than $\sim 180^\circ$ in extent are no longer
clumps, but radially confined rings that blend into the background disk.
Furthermore, the emission from such rings could be affected by the
subsequent collisional evolution of their constituents\footnote{The subsequent
collisional evolution of collisional fragments would not
affect their emission by much until $\theta > 180^\circ$, as the
collisional lifetime of this material, assuming it is the same as that
of the background disk (Fig.~\ref{fig:tccs}), is much longer than the age
of the clump (equation \ref{eq:tcl}).
This evolution would tend to reduce the clump's brightness, as collisions
remove the smallest dust first.}.
Here we define the \textbf{\textit{clump function}} to be the number of
clumps we expect to see in the disk that are smaller than $180^\circ$ in
extent and brighter than $F_\rmn{cl}$, $N(\theta<180^\circ,>F_\rmn{cl})$.
The number of clumps brighter than this limit, but with a maximum extent
smaller than $180^\circ$, can then be determined using the relation:
\begin{equation}
  N(<\theta,>F_\rmn{cl}) = N(<180^\circ,>F_\rmn{cl})\times
  (\theta/180^\circ). \label{eq:nthfcl}
\end{equation}

\begin{figure}
  \begin{center}
    \begin{tabular}{c}
       \psfig{figure=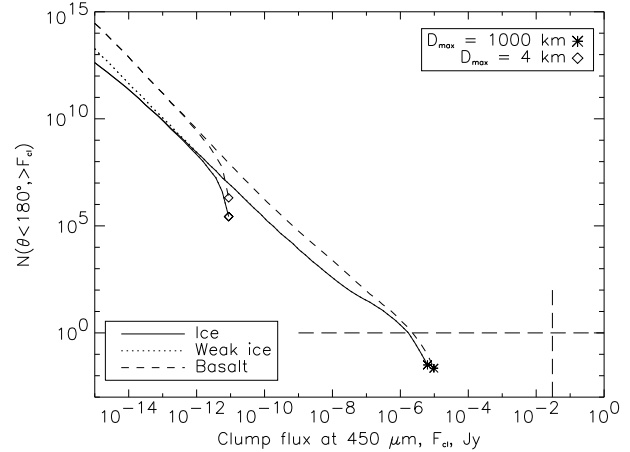,height=2.35in}
    \end{tabular}
    \caption{The \textit{clump function}, $N(\theta<180^\circ,>F_\rmn{cl})$, of the
    Fomalhaut disk, i.e., the number of clumps we expect to see that are brighter
    than $F_\rmn{cl}$ but smaller than $180^\circ$ in azimuthal extent.
    The solid, dotted and dashed lines correspond to the models for the
    collisional properties of \textbf{Ice}, \textbf{Weak Ice} and \textbf{Basalt}
    described in the text.
    The two lines for each model assume that the size distribution with
    $q_\rmn{d} = 1.84$ extends up to maximum planetesimal sizes of 4 km and
    1000 km.
    The \textit{intrinsic clumpiness} of the disk (the $F_\rmn{cl}$ at
    which $N(\theta<180^\circ,>F_\rmn{cl}) = 1$) is shown by the horizontal long
    dashed line.
    Its observed clumpiness, $F_\rmn{cl}=30$ mJy, is shown by the
    vertical long dashed line.
    For collisions to be the likely cause of this clumpiness we would expect
    the clump function to pass through the intersection of these lines.}
    \label{fig:fomngccl}
  \end{center}
\end{figure}

In Fig.~\ref{fig:fomngccl} we have plotted the clump function in Fomalhaut's
disk assuming that the disk is comprised of:
(i) just the planetesimals in the collisional cascade;
(ii) the collisional cascade plus an additional $\sim 1 M_\rmn{J}$ component of
primordial planetesimals with a size distribution defined by $q_\rmn{p} = 1.84$
extending up to those 1000 km in diameter.
The steepness of the clump function in this plot, combined with the linear
growth rate of clumps, implies that this clump function can be loosely
interpreted as the number of clumps we expect to see of flux
$\sim F_\rmn{cl}$ that are 90--$180^\circ$ in extent.
The origin of these clumps could be in any collision on the appropriate
line in Fig.~\ref{fig:fomccls}.
However, we find that clumps produced in the gravity regime are most
likely to be target debris\footnote{As target and impactor debris remains
in a clump for different lengths of time, we have treated the two
components separately.} from a threshold catastrophic collision;
e.g., any 1 $\mu$Jy clump we observe is most likely material created in
a collision between two bodies $\sim 300$ km in diameter.

We define the \textbf{\textit{intrinsic clumpiness}} of a disk as the flux
at which the clump function, $N(<180^\circ,>F_\rmn{cl}) = 1$.
This can be interpreted as the approximate flux of the largest single
clump we would expect to see in the disk.
However, as a disk contains a distribution of clump sizes, smaller
fainter clumps would also be observed down to a level determined by
the beam size.
Observations with a beam of width FWHM (in arcsec), would see all clumps
with $\theta < \theta_\rmn{FWHM}$, where $\theta_\rmn{FWHM} = 360^\circ
\times (\rmn{FWHM}\hspace{0.05in} r_\star)/(2\pi  r_\rmn{cl})$ ($r_\rmn{cl}$ is the
distance of the clump from the star in AU) as unresolved point sources;
those with $\theta_\rmn{FWHM} < \theta < 180^\circ$ would be resolved,
while larger ones would be rings rather than clumps.
Thus, the number of clumps we would expect to see as point sources above
a certain flux would be $\theta_\rmn{FWHM}/180^\circ$ times the clump
function, and the brightest unresolved point source we could expect to
see would be the flux at which the clump function,
$N(<180^\circ,>F_\rmn{cl}) = 180^\circ/\theta_\rmn{FWHM}$.

Fig.~\ref{fig:fomngccl} shows that if Fomalhaut's disk is comprised
solely of planetesimals in the collisional cascade, its largest (4 km)
planetesimals are so numerous, and their clumps last so
long, that at any one time we would expect to see roughly one million
clumps resulting from their destruction that are $<180^\circ$ in extent.
Thus even if the sensitivity of our observations was sufficient to
detect these clumps (which would emit just $\sim 10$ pJy), the disk
would appear smooth unless our beam size was also $< 60$ $\mu$arcsec!
However, if the disk also contains a primordial planetesimal distribution
as described in (ii) above, the disk's intrinsic clumpiness would be
$\sim 2$ $\mu$Jy (Fig.~\ref{fig:fomngccl}).
Brighter clumps could also exist in the disk (up to 10 $\mu$Jy),
however such large collisional events happen infrequently.
For these clumps, the clump function defines the probability of our
witnessing a clump that bright and smaller than $180^\circ$;
i.e., at any one time we have a 1:100 chance of seeing a 10 $\mu$Jy
clump (if the proposed distribution is correct).

Could intrinsic clumpiness be the likely cause of the clump
observed in Fomalhaut's disk (section 2)?
If so we would expect the derived intrinsic clumpiness at 450 $\mu$m
to be 30 mJy;
i.e., the clump function should pass through
$N(<180^\circ,>30$ mJy) $\approx 1$.
This is not true in either of the examples shown in
Fig.~\ref{fig:fomngccl}.
However, we have no reason to expect that $q_\rmn{p} = 1.84$,
nor that this distribution extends up to $D_\rmn{max} = 1000$ km.
A lower $q_\rmn{p}$ would increase the number of 1000 km
planetesimals in the disk, and so increase the number of clumps
expected from their destruction (although this would also imply a
disk more massive than $1M_J$, equation (\ref{eq:mp})).
The clump function could also be underestimated due to
uncertainties in our model.
Particularly, the lifetime of clumps produced by the destruction
of planetesimals in the gravity regime is not well known
(as discussed in section 6.3), and may have been
underestimated.
These factors could mean that the intrinsic clumpiness is actually
at the level of the break-up of the largest planetesimal in the
disk.
However, if $D_\rmn{max} = 1000$ km, this is just $\sim 10$ $\mu$Jy,
some 3000 times lower than that required for intrinsic clumpiness to
explain the observed clump.

However, it is possible that our model also underestimates the
amount of dust produced in collisions in the gravity regime, since
smaller dust grains might be predominant in the escaping fragments
from these collisions.
If the velocity imparted to the fragments is size dependent, and
smaller fragments attain higher velocities, then it would be the
largest fragments of the original distribution that reaccumulate,
while the smaller ones are more likely to be released.
Furthermore, the surfaces of large planetesimals are expected to
be covered with large quantities of small dust grains in the form
of a regolith (see section 7.2);
these grains would be preferentially released in a collision.
It has not been possible to include such effects in our model
because the outcome for $\mu$m-sized dust from km-scale impacts
has yet to be studied.
Based on arguments outlined in section 7.1, the most our model
could be underestimating the flux by is a factor of
1000\footnote{The collisional model predicts that the destruction
of a 1000 km planetesimal would result in dust that emits a flux
of 10 $\mu$Jy, whereas the maximum possible emission from a pulverised
1000 km body, i.e., when it is ground into 150 $\mu$m--1 mm dust,
is 10 mJy (Fig.~\ref{fig:fommcl}b)}.
However, the real value is likely to be much less than this.
In any case, collisions among non-gravitationally focussing
planetesimals result in insufficient flux for them to be the cause
of the observed clump.

\subsection{For Future Consideration}
Could Fomalhaut's intrinsic clumpiness be observed with future
instrumentation?
With a beam size of 0.6 arcsec, Fig.~\ref{fig:fomngccl} implies that
we could expect to observe clumps in Fomalhaut's disk between
$\sim 0.1$ $\mu$Jy (unresolved point sources) and 2 $\mu$Jy
(resolved clumps) at 450 $\mu$m.
Depending on the amount of $\mu$m-sized dust produced in collisions,
these limits correspond to clumps of between 1 $\mu$Jy and
0.02 mJy at 25 $\mu$m (Fig.~\ref{fig:fomsedcldmin}).
A 0.6 arcsec beam is close to that of the \textit{ALMA} in compact
configuration at 450 $\mu$m, or that of a mid-IR instrument operating
at 25 $\mu$m on an 8 m telescope.
While the clumpiness of Fomalhaut's disk proposed above is at
least an order of magnitude below the anticipated sensitivities
of such instruments, the above figures include uncertainties
(such as the potential factor of 1000 underestimate of the amount of
dust created in a collision, section 6.4) which could render these
clumps detectable.
If so, their observed magnitude and size distribution could be used to
constrain the primordial planetesimal population.

Also, we note that this type of clumpiness could provide confusion
for a mission to find terrestrial planets using space-based 10 $\mu$m
interferometry, such as the designs proposed for \textit{DARWIN} or
\textit{TPF} (e.g., Mennesson \& Mariotti 1997).
The biggest potential impact of disk clumpiness on such a mission would
come from a warm zodiacal cloud-like disk $\sim 3$ AU from the star;
this topic is discussed in Wyatt (2001).
However, here we point out that if viewed at the distance of Fomalhaut, the
Earth would have a 25 $\mu$m flux of $\sim 0.8$ $\mu$Jy, while
the primordial planetesimal distribution proposed in Fig.~\ref{fig:fomngccl}
implies that the Fomalhaut disk could contain a comparable $\sim 0.2$
$\mu$Jy clump that would be unresolved in a 0.05 arcsec beam, which would
be that of a 100 m baseline interferometer at this wavelength.

\section{The Collisional Clump Hypothesis}
Consider the magnitude of collision required to produce the clump observed
in the Fomalhaut disk.
Section 6.4 showed that this clump cannot have been produced by the intrinsic
clumpiness of the collisional cascade, or by the break-up of the primordial
planetesimal population, because such collisions do not produce enough dust.
Thus we start by determining the smallest planetesimal that could have
produced the clump using a simple model in which the clump is comprised of
grains of just one size.

\subsection{Single size dust grained clump}
\begin{figure}
  \begin{center}
    \begin{tabular}{rl}
       \textbf{(a)} & \hspace{-0.3in} \psfig{figure=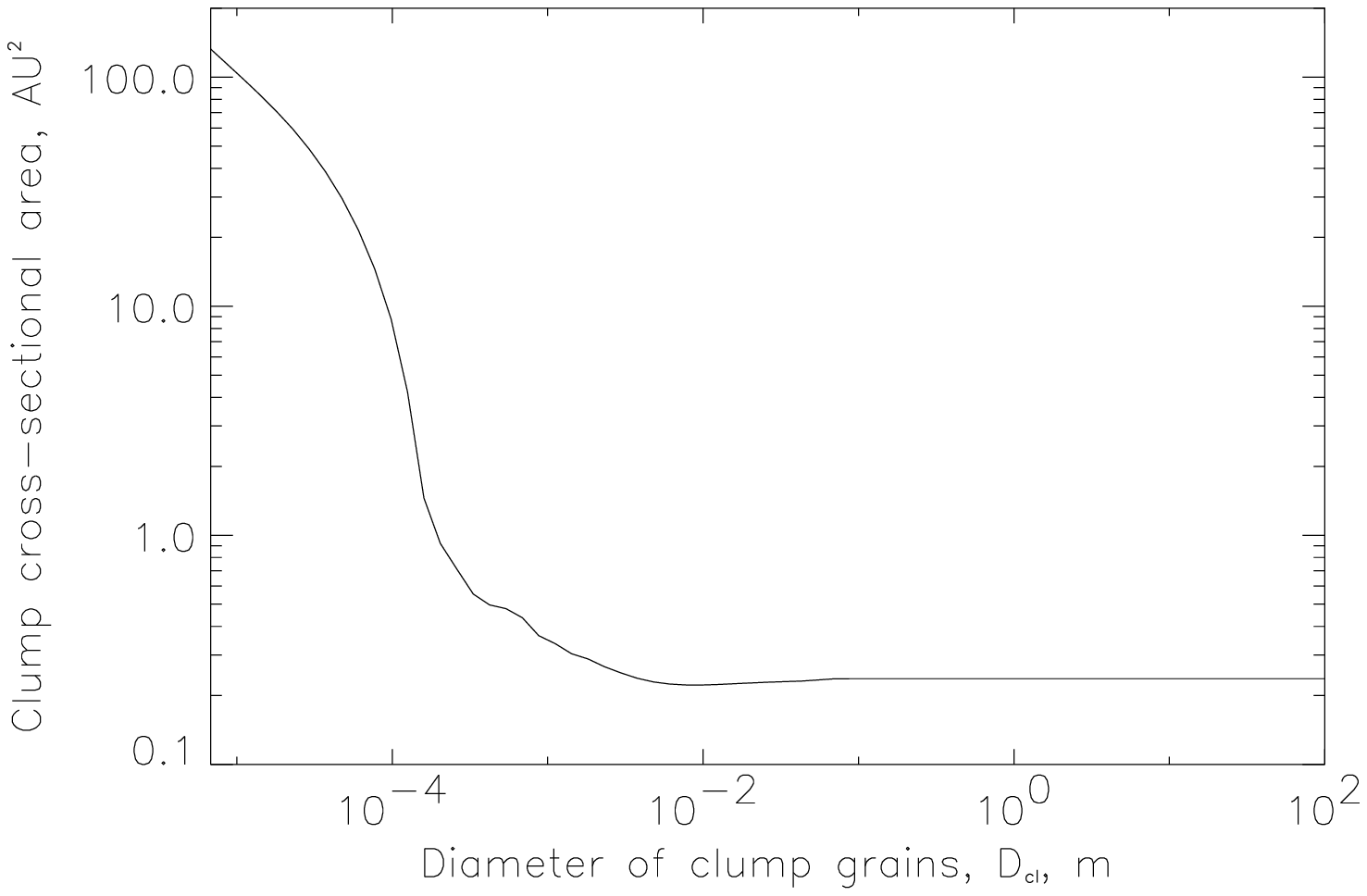,height=1.9in} \\[0.15in]
       \textbf{(b)} & \hspace{-0.4in} \psfig{figure=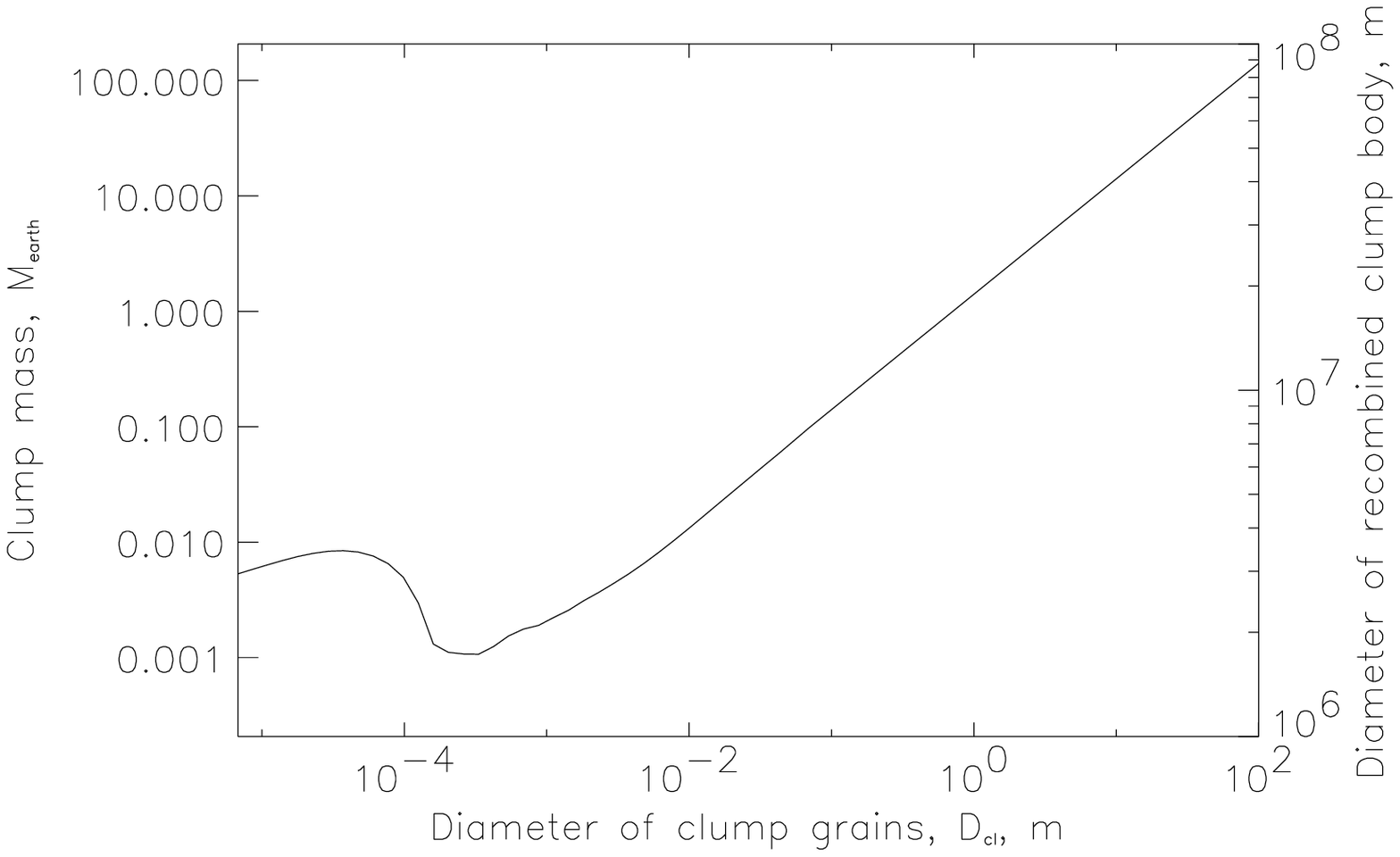,height=1.98in} \\[0.25in]
       \textbf{(c)} & \hspace{-0.45in} \psfig{figure=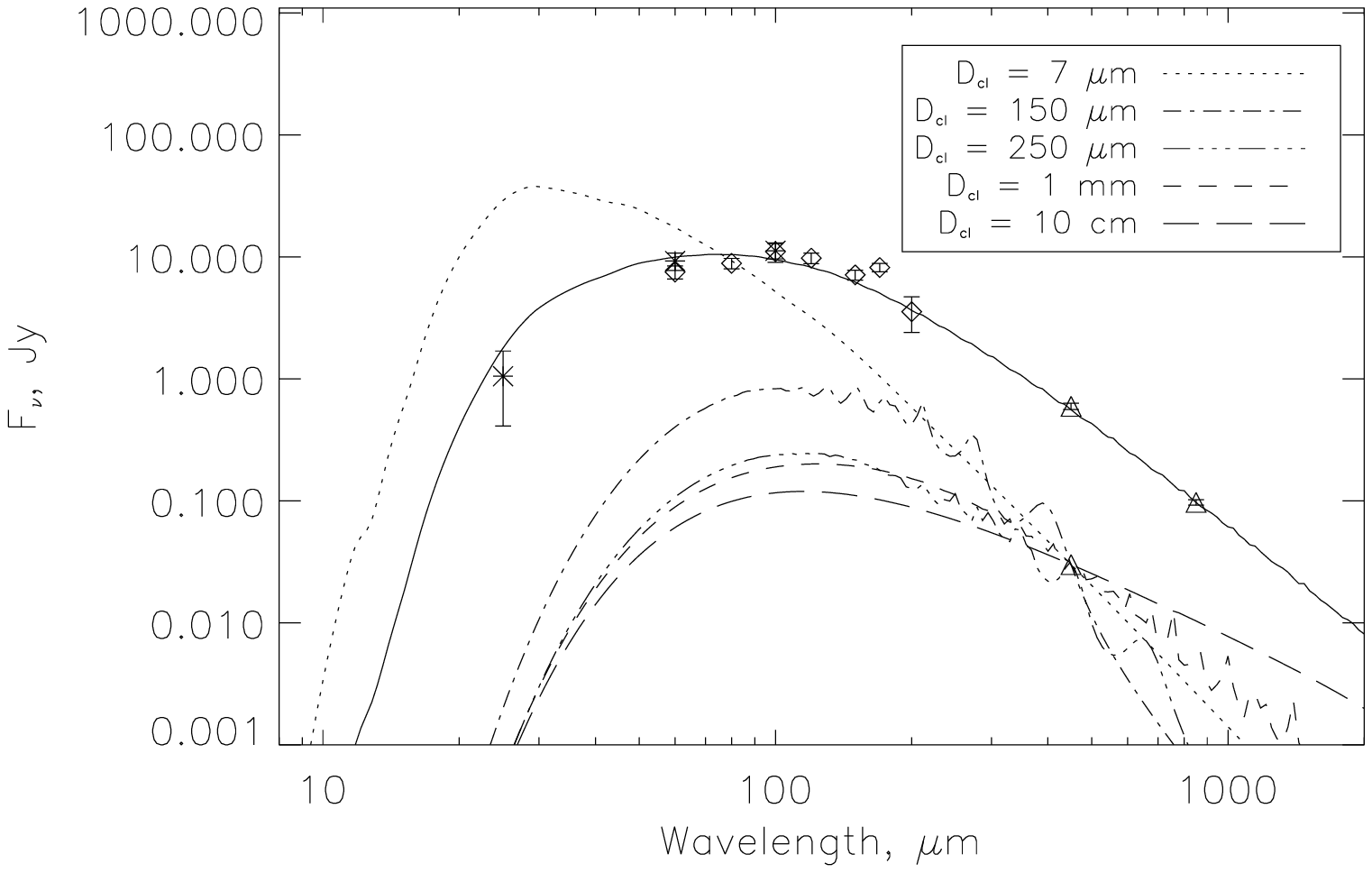,height=1.88in} \\
    \end{tabular}
    \caption{The total cross-sectional area \textbf{(a)} and mass
    \textbf{(b)} of material in Fomalhaut's clump, as well as this clump's
    SED \textbf{(c)}, assuming it is comprised of grains of just one size,
    $D_\rmn{cl}$, and setting the clump's emission to 30 mJy at 450 $\mu$m.
    As in section 6, the clump is assumed to be composed of the same material
    as that inferred for the smooth ring and the clump's emission is calculated
    assuming that this material is spread across the same range of distances
    from the star as the smooth ring.
    Also shown in \textbf{(b)} is the diameter of the body that the
    clump mass would correspond to when recombined into a single body,
    and \textbf{(c)} also shows the observed SED (symbols) and the fit to
    this SED assuming all this emission comes from the smooth ring (solid
    line).}
    \label{fig:fommcl}
  \end{center}
\end{figure}

If the dust grains in the clump are all of the same size, $D_\rmn{cl}$,
then Fig.~\ref{fig:fommcl}a shows how the observed 450 $\mu$m flux of
the clump, $F_\nu (450$ $\mu$m) = 30 mJy, determines the total
cross-sectional area of material in the clump, $\sigma_\rmn{cl}$
(found by inverting equation (\ref{eq:fnu})), for different $D_\rmn{cl}$.
Thus the projected area of this clump must be at least
0.2 AU$^2$, much larger if a significant fraction of the
material is in dust smaller than 1 mm.
Fig.~\ref{fig:fommcl}b also shows how this flux determines the
total mass of material in the clump, where
$m_\rmn{cl} = (2/3)\rho \sigma_\rmn{cl}D_\rmn{cl}$.
Thus the clump must have a mass of at least a thousandth that of the
Earth, and would only be this small if all the grains in the clump
are between 150 $\mu$m and 1 mm in diameter.
If there is a significant fraction of mass in particles smaller than
150 $\mu$m then we would infer a more massive clump, since these particles
do not emit efficiently at 450 $\mu$m.
Similarly we would infer a more massive clump if there is a significant
fraction of mass in particles larger than 1 mm.
It would be possible to constrain the size of particles in the clump
if the clump could be imaged at shorter wavelengths (see
Fig.~\ref{fig:fommcl}c).
However, for now we can rule out the possibility that the clump consists
of grains that are all smaller than $\sim 100$ $\mu$m, since the
emission from such a clump would exceed that observed, presumably
from the smooth ring, at mid- to far-IR wavelengths.

Thus if the material in this clump originated from a single body, that
planetesimal would have to have been at least 1400 km in diameter.
To understand how such a large body, which must have formed through
runaway growth, and for which gravitational focussing is important,
interacts with the planetesimal disk, we start by considering how such
a planetesimal grew in the first place.

\subsection{Runaway planetesimal growth}
The runaway population are the largest members of the distribution of
planetesimals growing in the protostellar nebula.
They grew initially in the same way as the rest of the planetesimals, 
by the accretion of nearby, usually much smaller planetesimals.
However, kinetic energy exchange in encounters with other planetesimals
(i.e., dynamical friction) led to larger planetesimals having more circular
orbits (Wetherill \& Stewart 1993).
The consequent low relative velocity and high gravitational focussing factor
for collisions with large bodies (e.g., equation (\ref{eq:fddim})),
meant that large bodies (once larger than $\sim 1$ km) grew more rapidly
than smaller planetesimals, creating a bimodal size distribution --
an accumulation cascade and a population of large, so-called runaway,
planetesimals.
This runaway growth would have been relatively short, typically several
Myr (e.g., Kenyon 2002), and would have been halted once the runaways
became large enough to perturb the disk just enough to affect their
collisional velocities (i.e., once larger than a few hundred km).
The runaways then continue to grow, but at a slower rate than smaller
planetesimals, so-called oligarchic growth (e.g., Kokubo \& Ida 2000).
Once these runaways are large enough, however, their gravitational
perturbations increase the impact velocities of members of the
accumulation cascade so that collisions result in their net
destruction, rather than in their net accretion.

During this oligarchic growth, the runaways try to remain dynamically
isolated from each other.
However, this is only possible as long as their separation is at
least 5 Hill's radii (Kokubo \& Ida 1995),
where a planetesimal's Hill's radius is $r_\rmn{H} =
a(M_\rmn{pl}/3M_\star)^{1/3}$.
A runaway's increasing mass means that any isolation is not permanent
and interactions with other runaways are inevitable.
These interactions would result in both collisions amongst the runaway
population, as well as increases in the eccentricities of this population,
thus increasing their feeding zones (e.g., Weidenschilling et al.~1997)
and making mutual collisions more likely.
Their growth would eventually be halted once they had scattered or
accreted all of the planetesimals in their feeding zone.
The angular momentum exchange caused by this scattering could cause
the orbit of the newly formed planet to migrate either towards or
away from the star (e.g., Hahn \& Malhotra 1999).

A runaway body would have started as an incoherent body built up of
layers of regolith.
Mutual collisions between runaways would have resulted in the runaway often
being completely fragmented, then reaccumulating.
Since smaller fragments attain higher ejection velocities, the reaccumulated
body would be sorted according to grain size with the larger grains towards
the centre and fine grains on the surface (e.g., Britt \& Consolmagno 2001).
Impacts of smaller bodies onto the runaway would result in the comminution
of the runaway's surface regolith layer (e.g., McKay, Swindle \&
Greenberg 1989).
Planetesimals in the solar system have such regoliths:
the Moon's regolith is 5-10 m deep in the mare region,
and is comprised mostly (80--90 \%) of dust grains smaller than 1 mm in size
(e.g., Heiken 1975), while that of 20 km diameter Phobos is thought to be
more than 100 m thick in places (Veverka \& Thomas 1979), and
the regolith layer on 200 km-sized asteroids is predicted to extend to a
few km in depth (Housen \& Wilkening 1982).
Thus we expect runaways to be composed of a megaregolith structure on
top of which is a deep dusty regolith layer.

\subsection{Observing the final throes of planet formation?}
Within this scenario there are three ways in which the observed
dust clump could be associated with a runaway planetesimal in the ring:
\textbf{(i)} it is unbound material generated in a recent collision between
two runaways;
\textbf{(ii)} it is bound material constantly replenished by impacts into
a growing runaway;
\textbf{(iii)} it is an infalling envelope of dust that is being accreted
from the collisional cascade onto the runaway.
The clump could also be associated with a planet orbiting interior to
the ring:
\textbf{(iv)} its dust could have been created by the break-up of
planetesimals that are trapped in resonance with the planet.

\subsubsection{Unbound runaway-runaway collisional clump}
If we assume that the $M_\oplus/1000$ of dust that we see originated in
just one collision, then it must have originated from a parent
planetesimal more than 1400 km in diameter.
For this dust to be released from the gravity of the parent body,
the impacting planetesimal must have had enough mass that the kinetic
energy imparted to the fragments exceeded the escape velocity.
Since impacts with gravitationally focussing planetesimals take place
at the escape velocity, at best an impactor can only release
$f_\rmn{KE}$ times its own mass.
Thus the observed clump, if the product of one collision, must have
involved the collision of two runaway planetesimals at least 1400 km
in diameter.
In fact, the planetesimals may have to have been significantly larger
than this, since experiments show that transmission of kinetic energy is
much less efficient for collisions into sand than into rock (e.g.,
Holsapple 1993) leading some authors to speculate that
$f_\rmn{KE} \approx 0.01$ for collisions with such rubble piles
(Campo Bagatin et al.~2001).

How likely are we to be observing a collision between two runaways?
Planet formation simulations do predict the formation of a significant
population of runaways.
Also, we know that runaway-runaway collisions do occur:
based on the similarity of the abundances of isotopes of oxygen in
lunar samples to those on the Earth, Wiechert et al.~(2001) concluded
that the Mars-sized impactor that collided with the Earth to form the
Moon formed at the same heliocentric distance as the Earth (i.e., it
was a nearby runaway).
To have grown to $>1400$ km, the runaways in Fomalhaut's disk must
have grown in isolation from each other, with at least 5$r_\rmn{H}$
separation.
This means that there can be at most one hundred $M_\oplus/1000$
runaways in Fomalhaut's 50 AU wide ring.
Since each collision would result in the merger of the runaways
and so in the loss of one of the runaways, this leaves at most
about 100 possible collisions resulting in the observed clump,
that is if the runaways are not scattered out of the disk by their
mutual perturbations before they collide.

The length of time this material would remain in a clump depends
on its ejection velocity.
The only studies that can throw light on the outcome of such
collisions are the SPH simulations of the formation of the
Moon (e.g., Cameron 1997; Canup \& Asphaug 2001) and of the
asteroid families (Michel et al.~2001).
However, even these do not predict the outcome for collision
fragments that are smaller than their resolution ($>$ km), or
indeed what would happen to the dusty regoliths on the colliding
bodies.
What they do show, however, is that in a collision between two
runaways a significant mass of fragments does escape their
gravitational field (see table 1 of Canup \& Asphaug 2001).
Thus such collisions would result in the formation of an unbound
clump.
Also the ejection velocities of fragments from collisions between
massive, but non-gravitationally focussing, planetesimals imply a
clump lifetime (equation (\ref{eq:tcl})) of 20-30 orbital periods
for collisions in the asteroid belt (Michel et al.~2001).
Thus if this clump lifetime is also typical for runaway-runaway
collisions in Fomalhaut's disk, we could only expect to see a
clump from their mutual collisions for a total of $\sim 3$ Myr
of the disk's life.

This is an extremely short window to be witnessing the product of
runaway-runaway collisions given the age of the system (3 Myr is
1--2 per cent of the disk's life).
However the lifetime of a clump created in a collision between
gravitationally focussing bodies in Fomalhaut's disk could be an order
of magnitude longer than those of clumps created in the asteroid belt.
Also planet formation models predict that the growth of the runaways
to sizes large enough both to cause the observed clump and to
ignite of the collisional cascade can take a few hundred Myr at
the distance of Fomalhaut's disk (e.g., Kenyon \& Bromley 2001).
Thus we leave this topic to be explored using more detailed planet
formation and collision models, but remark that it appears very
unlikely at this stage that runaway-runaway collisions are the cause
of the observed clump.

\subsubsection{Bound collisionally replenished envelope (CRE)}
Impacts into a runaway planetesimal would launch a cloud of regolith
dust from its surface.
The majority of this dust would then collapse back onto the runaway.
The maximum distance out to which dust would remain bound to the runaway
would be to its Hill's radius, since outside this its gravity would be
weaker than that of the star;
this corresponds to $\sim 1.2$ AU for a $1M_\oplus$ planetesimal
orbiting at 150 AU around Fomalhaut.
Thus for the observed dust clump, which has a projected area of
$> 0.2$ AU$^2$ (section 7.1), to be bound to its parent
planetesimal, that planetesimal would have to have a mass of at
least $0.01M_\oplus$.
Here we consider whether Fomalhaut's clump could be
caused by repeated collisions onto a runaway from non-gravitationally
focussing planetesimals sustaining a dusty envelope of
regolith material extending to the Hill's radius.
Since all collisions with runaways occur at the escape
velocity, and the ejection velocity required for dust to reach to
the Hill's radius is close to (just $5 \times 10^{-5}$
times less than) that escape velocity, energy constraints imply
that the maximum amount of regolith dust that an impacting
planetesimal could launch to such a distance would be equal to
its mass.
Thus the maximum mass of material in such an envelope at any one time
is equal to the mass accretion rate onto the runaway times the maximum
length of time that the bound dust could remain in the envelope.

\begin{figure}
  \begin{center}
    \begin{tabular}{rl}
       \textbf{(a)} & \hspace{-0.3in} \psfig{figure=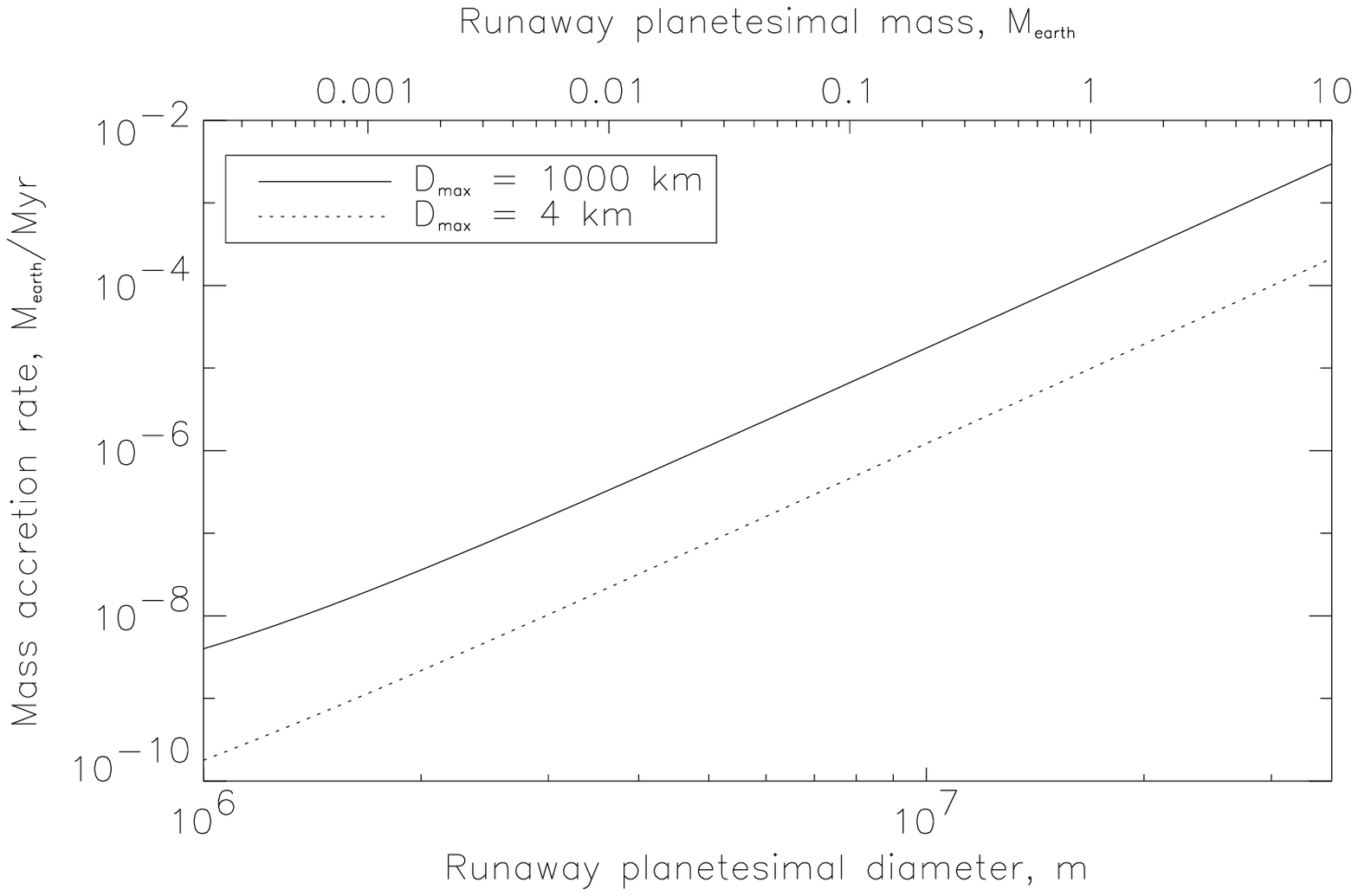,height=2.1in} \\[0.1in]
       \textbf{(b)} & \hspace{-0.3in} \psfig{figure=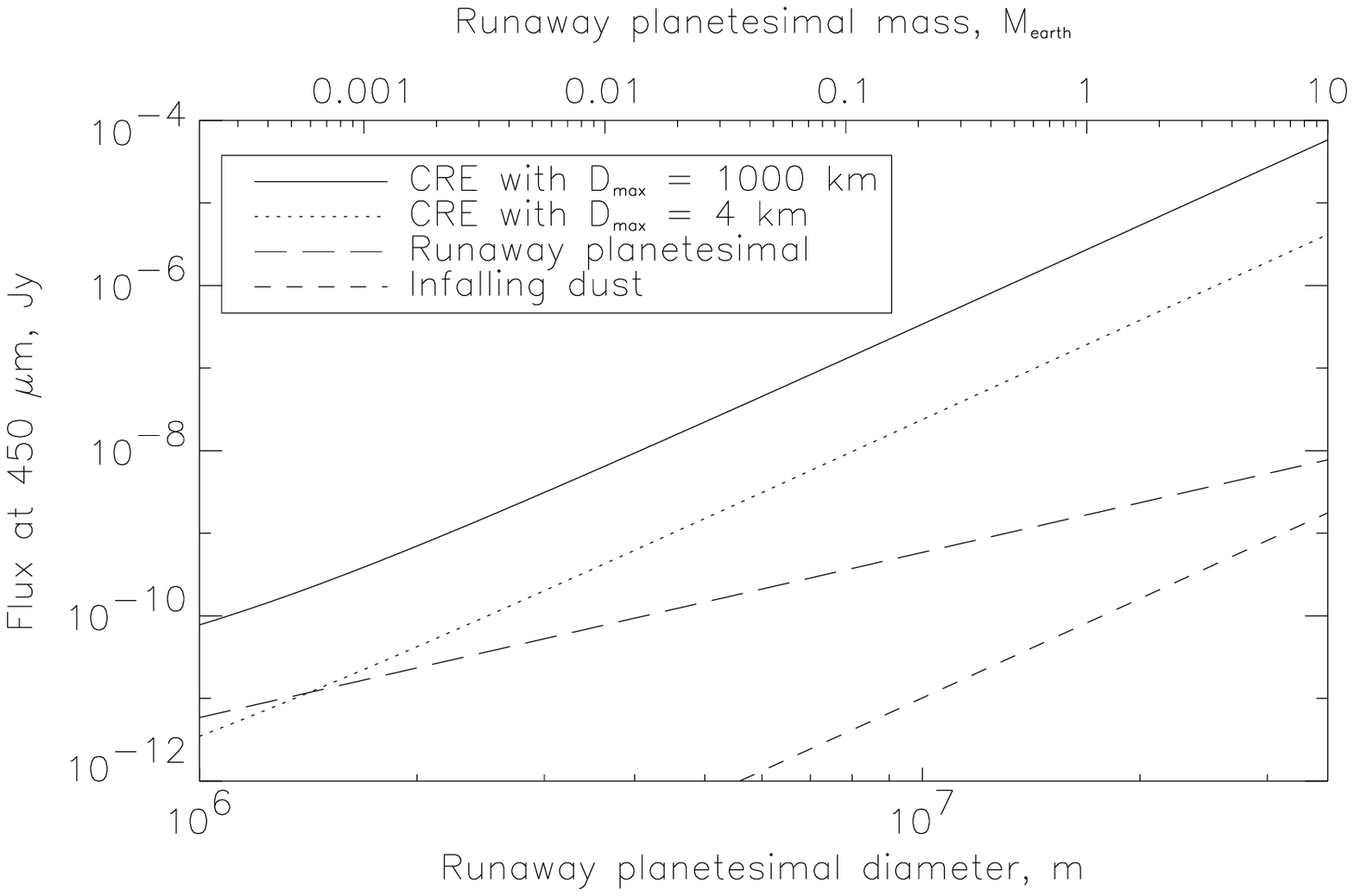,height=2.1in} \\[0.1in]
    \end{tabular}
    \caption{The interaction of a runaway planetesimal with the non-runaway
    population in Fomalhaut's disk:
    \textbf{(a)} the rate of accretion of mass onto the runaway;
    \textbf{(b)} the flux of clump expected from this interaction due to
    both the collisionally replenished envelope (CRE) of regolith material
    and the infalling dust envelope, as well as the emission from
    the runaway itself (assuming a temperature of $\sim 45$ K).
    Since a runaway accretes most of its mass from the largest members of the
    population, the mass accretion rate and the flux from the CRE
    are shown for both $D_\rmn{max} = $ 4 km and 1000 km (dotted and
    solid lines respectively).}
    \label{fig:fomfrun}
  \end{center}
\end{figure}

The mass accretion rate of a runaway of size $D_\rmn{run}$ is
given by
\begin{equation}
  \dot{M}_\rmn{accr}(D_\rmn{run}) = \int_{D_\rmn{min}}^{D_\rmn{max}}
    (\pi \rho /6)D_\rmn{im}^3
    R_\rmn{col}(D_\rmn{run},D_\rmn{im})
    \rmn{d}D_\rmn{im}.
  \label{eq:maccr}
\end{equation}
This is plotted in Fig.~\ref{fig:fomfrun}a for size distributions that
extend up to 4 km (i.e., the collisional cascade) and 1000 km (i.e.,
including a hypothetical primordial planetesimal population).
Note that since the growth rate of runaways in the current disk is
relatively slow, they must have formed when the eccentricities and
inclinations of disk material were much lower, since then the accretion
rate would have been higher ($\dot{M}_\rmn{accr} \propto [If(e,I)]^{-1}$).
The maximum amount of time dust from any one collision would remain
in the envelope would be twice the free-fall timescale from the Hill's
radius (e.g., Love \& Ahrens 1996), i.e., about half an orbital period,
or 650 years at 150 AU from Fomalhaut.
Some material could, however, remain for longer if injected into orbit
around the runaway (e.g., Canup \& Asphaug 2001; Michel et al.~2001),

The maximum 450 $\mu$m flux a given mass of material in Fomalhaut's disk
could emit can be calculated from Fig.~\ref{fig:fommcl}b.
If all this mass was in dust 150 $\mu$m to 1 mm in size, this
would emit 30 mJy/$0.001M_\oplus$.
Fig.~\ref{fig:fomfrun}b uses this to show the maximum flux of such
a collisionally replenished envelope (CRE).
This shows that even the CREs surrounding $10M_\oplus$ planets could
not explain the observed dust clump.
However, the flux of these CREs could surpass that of the runaway
itself by up to a factor of 10,000 for the 450 $\mu$m flux of a
10 $M_\oplus$ planet (or perhaps even higher at shorter wavelengths,
see Fig.~\ref{fig:fommcl}c).
These CREs could also be bright enough that it is the distribution
of massive runaways that determines a disk's intrinsic clumpiness,
rather than the distribution of non-gravitationally focussing
planetesimals (section 6.4).

\subsubsection{Infalling dust envelope}
Another reason why there would be an enhanced density of dust in the
vicinity of a runaway is that it is accreting dust from the disk's
collisional cascade.
The total cross-sectional area of material in this infalling envelope
is approximately the runaway's rate of accretion of cross-sectional
area, given by
\begin{equation}
  \dot{\sigma}_\rmn{accr}(D_\rmn{run}) = \int_{D_\rmn{min}}^{D_\rmn{max}}
    (\pi /4)D_\rmn{im}^2
    R_\rmn{col}(D_\rmn{run},D_\rmn{im})
    \rmn{d}D_\rmn{im},
  \label{eq:saccr}
\end{equation}
multiplied by the infall timescale, which is the freefall timescale
from its Hill's radius (one quarter of an orbital period).
The flux from this infalling material can then be calculated by considering
that the disk emission of 0.595 Jy at 450 $\mu$m (Table 1) is predicted to
originate from $\sigma_\rmn{tot} = 33.7$ AU$^2$ of material (section 4.3).
The resulting flux is plotted in Fig.~\ref{fig:fomfrun}b, which shows that
such an infalling envelope can be ignored as it is fainter even than the
runaway.

\subsubsection{Dusty resonant ring}
While no planets have been detected around Fomalhaut,
the existence of a planetary system orbiting within the observed dust
ring would naturally explain the lack of dust emission there, since
such a system would have cleared the planetesimal disk within which
it formed.
As stated in section 7.2, this clearing could have caused the
planets to migrate outwards.
In the course of this migration, the outer mean motion resonances
of the outermost planet of the system would have swept through
the disk of planetesimals outside the planet's orbit.
Many of these planetesimals would have become trapped in these
resonances and would then have migrated out with the planet,
while those that did not get trapped would have been scattered
by the approaching planet.
This is the mechanism which is invoked to explain the large
number of Plutinos, Kuiper belt objects that are trapped in
Neptune's 2:3 resonance (e.g., Jewitt 1999).
Hahn \& Malhotra (1999) estimated that the required migration of
Neptune's orbit from 23 AU to 30 AU could have been caused by the
clearing of a 50 $M_\oplus$ planetesimal disk over the first 50 Myr
of the solar system.
We note that these last two figures are comparable to the inferred
mass and age of Fomalhaut's disk, thus it could be argued that
similar migration and trapping are to be expected in this system
had a planet formed in a (now scattered) planetesimal disk
interior to the one we see today.

The paths of eccentric resonant orbits when plotted in a frame
co-rotating with the planet's mean motion exhibit loops which occur at the
pericentres of the orbits (see fig.~8.4c of Murray \& Dermott 1999).
As only those resonant orbits that have their conjunction with the
planet when at their apocentres are stable (at least when the planet
has a low eccentricity), these loops only occur at specific locations
relative to the planet.
Planetesimals that are evenly distributed within these stable resonances
would be most densely concentrated in these loop regions.
Thus, assuming that the fragments created by the collisional break-up of
such resonant planetesimals also remain in resonance, the resulting dust
ring would be clumpy (Wyatt 1999).
The number and distribution of these clumps would be determined
by the fraction of planetesimals that are trapped in the different
resonances.
The one clump in Fomalhaut's ring implies that the material
is trapped in a 1:2 resonance, although other clumps could be
hidden in the lobes.
Very few Kuiper belt objects are observed to be trapped in
Neptune's 1:2 resonance, however this may be an observational
limit, or because the primordial Kuiper belt did not extend
out that far, and/or Neptune's orbit did not migrate far enough.
The fraction of planetesimals that we require to be trapped in the
the 1:2 resonance to cause the observed 30 mJy clump would be of the
order of 5 per cent, since resonant material can spend a significant
fraction of its time in the clump region (the actual fraction is
determined by its eccentricity) and while in the clump this material
would be closer to the star and so hotter and brighter than that in
the rest of the ring.

If this interpretation is correct we would predict that the
planet orbits at 80 AU with a period of about 500 years, and
that it is currently located on the opposite side of the star
from the dust clump.
We would also predict that the clump would orbit the star
with the planet (direction unknown) causing motion on the
sky of up to 0.2 arcsec/year, and that the clump would be
hotter than the rest of the ring.
A model for this resonant structure, and the constraints it sets
upon the planet's mass, and that of the original planetesimal disk,
is beyond the scope of this paper and will be discussed in a
forthcoming paper.

We note that the resonant ring we propose here is formed
in a different manner to that proposed to explain the Vega and
$\epsilon$ Eridani dust clumps (Ozernoy et al.~2000; Wilner et
al.~2002).
The dust in their rings is trapped into planetary resonances
through inward dust migration caused by P-R drag, rather
than by outward planetary migration.
Thus their rings are akin to the Earth's resonant ring
(Dermott et al.~1994).
We contend that this mechanism would be inefficient in
this system, as grains would be destroyed by collisions before
they migrate into the resonances by P-R drag (e.g., section 3.3;
WDT99; Wyatt 1999).

\section{Conclusions}
We have presented a detailed study of collisional processes
across the whole size range in material in the Fomalhaut debris
disk.
This is the first such study of a debris disk that we are aware
of.
We have shown that the spatial distribution of the dust derived
from the 450 $\mu$m image of the disk, combined with modelling
of its SED, are sufficient to constrain the dust's size distribution
but not to set stringent constraints on its composition.
The size distribution we inferred is the same as that expected
from a disk in collisional equilibrium, although we recognize that
our parameterization of this distribution is only an approximation.
The emission that we observe originates from material in the size
range 7 $\mu$m--0.2 m, however collisional lifetime arguments imply
that this material originated in a cascade involving the break-up
of planetesimals up to 4 km in size.
The inferred mass of material in this collisional cascade
is $20-30 M_\oplus$.
We speculate that there may be a population of primordial
planetesimals with sizes between 4-1000 km, however these
collide too infrequently for their collisional debris to
have contributed to the dust we observe today.
The slope of the size distribution of these planetesimals
could be similar to that of the cascade and extend up to 1000
km based on models of planet formation.
Dust clumps caused by the break-up of these planetesimals
would be at too low a level to be detected with current
technology, however it may be possible to observe these
clumps in the future (e.g., with the
\textit{ALMA}).
With such observations we would be able to constrain the
population of these large planetesimals --- indeed
this may be the only way of detecting their presence.
There may also be a population of runaway planetesimals
larger than 1000 km which would still be growing.
It may be possible to detect these directly, since they
would always be enveloped by a cloud of dust launched
from their regolith by repeated collisions.
These dust envelopes could also be at a level which it
may be possible to observe in the future.

We propose two possible origins for the 30 mJy clump that
is observed in Fomalhaut's disk.
First it could have been created in a collision between two
runaway planetesimals.
We are limited in being able to assess the likelihood of
this by uncertainties in the outcome of collisions between
such massive bodies, as well as by uncertainties in the number
and collision frequencies of such bodies at the end of planet
formation.
Our best estimate, however, shows that this possibility is
unlikely unless both the formation of the runaways and the
ignition of the collision cascade occurred within the last
few Myr.
Another possibility is that $\sim$ 5 per cent of the
planetesimals in the ring were trapped in 1:2 resonance
with an inner planet when it migrated out due
to the clearing of a residual planetesimal disk.
This hypothesis is strengthened by the fact that the mass and
age of Fomalhaut's disk are comparable to that required for
Neptune's migration and the consequent resonant trapping of
Kuiper belt objects.
Constraints on the mass of Fomalhaut's hypothetical planet
will be addressed in a future paper.
We predict that the orbital motion of such a resonant clump
(0.2 arcsec/year) could be detectable within a few years,
and that this clump would be more prominent at shorter
wavelengths.

It is clear from this study that collisions are fundamental
to the formation and evolution of debris disks, not only
as the mechanism which creates the dust we see, but as a cause
of potentially observable structure in the disk.
Unless future collisional and planet formation models can
prove otherwise, it appears unlikely that Fomalhaut's
observed clump is collisional in origin.
However, this does not rule out a collisional origin for
the clumps seen in other debris disks.
These disks will be the subject of a future paper.


\end{document}